\documentclass[letterpaper,twocolumn,amsmath,amssymb,aps,pra,floatfix,showkeys,footinbib,superscriptaddress,preprintnumbers,10pt]{revtex4-2}
\newcommand{\comment}[1]{}

\usepackage{graphicx}
\usepackage{dcolumn}
\usepackage{bm}

\usepackage{orcidlink}

\usepackage{hyperref}

\newcommand{\JQI}{Joint Quantum Institute, National Institute of Standards and Technology and University of Maryland, College Park, MD, USA}

\begin{document}

\title{BIFROST: A First-Principles Model of Polarization Mode Dispersion in Optical Fiber}

\author{Patrick R. Banner\,\orcidlink{0009-0006-9957-4996}}
\email{pbanner@terpmail.umd.edu}
\affiliation{\JQI}
\author{S. L. Rolston\,\orcidlink{0000-0003-1671-4190}}
\affiliation{\JQI}
\author{Joseph W. Britton\,\orcidlink{0000-0001-8103-7347}}
\affiliation{Department of Physics, University of Maryland, College Park, MD, USA}
\affiliation{CCDC Army Research Laboratory, Adelphi, MD, USA}
\affiliation{Quantum Technology Center, University of Maryland, College Park, MD, USA}

\date{\today}

\begin{abstract}
We present BIFROST, a first-principles model of polarization mode dispersion (PMD) in optical fibers. Unlike conventional models, BIFROST employs physically motivated representations of the PMD properties of fibers, allowing users to computationally investigate real-world fibers in ways that are connected to physical parameters such as environmental temperature and external stresses. Our model, implemented in an open-source Python module, incorporates birefringence from core geometry, material properties, environmental stress, and fiber spinning. We validate our model by examining commercial fiber specifications, fiber-paddle measurements, and published PMD statistics for deployed fiber links, and we showcase BIFROST's predictive power by considering wavelength-division-multiplexed PMD compensation schemes for polarization-encoded quantum networks. BIFROST's physical grounding enables investigations into such questions as the sensitivity of fiber sensors, the evaluation of PMD mitigation strategies in quantum networks, and many more applications across fiber technologies.
\end{abstract}


\maketitle

\section{Introduction}

Optical fiber has revolutionized the modern world. It delivers communications signals across neighborhoods, cities, continents, and oceans; illuminates and images hard-to-reach places in medicine and industry \cite{OCT_PMD, Buildings1, Buildings2}; senses temperature, strain, pressure, vibration, and rotation \cite{FiberSensingReviewOld, FBGSensorReview, ScatteringFiberSensorReviewAPL, Mecozzi}; synchronizes atomic clocks \cite{AtomicClocksReview}; and transmits quantum information \cite{QuantumNetworkFibers_RMP, QunnectEntanglement, PMDMeas3_Quantum}. 

In many of these applications, the polarization of the guided light is a critical consideration. Due to birefringences in the fiber, the polarization varies in time, a phenomenon known as polarization mode dispersion (PMD) \cite{Rashleigh, PMD_PNASReview}. PMD is an enabling phenomenon for some applications and is detrimental for others. In interferometric fiber sensors \cite{FiberSensingReviewOld, FiberInterferometers, FiberGyroscope}, PMD causes polarization fading \cite{PolFadingReview}, decreasing the sensitivity, while in fiber Bragg grating sensors \cite{FiberTempSensors}, PMD variation causes drifts in the resonance condition \cite{FBGSensor_PMDBad}. Conversely, birefringence enables sensing of directional forces \cite{FBGSensorReview, FBGSensor1, ScatteringFiberSensorReviewAPL, Mecozzi}. Additionally, interferometry techniques to improve the resolution and sensitivity of optical telescopes involve fiber polarization properties \cite{LongBaseline1, InterferometryAstro}. Finally, in the context of quantum networks \cite{QNReview_RMP_Azuma}, addressing the limits due to fiber birefringence is an active area of research \cite{QunnectEntanglement, PMDMeas_Quantum1, PMDMeas2_Quantum, PMDComp_QKD1, PMDComp_QKD2, PMDComp_QKD3, PMD_TDM1, PMD_TDM2, PMD_TDM3, PMD_TDM4, PMD_TDM5, PMD_TDM6, PMD_TDM7, PMD_TDM8, PMD_TDM9, PMD_WDM1, PMD_WDM2, PMD_WDM3, PMD_WDM4}.

Whether PMD is a challenge to be mitigated or a tool to be used, investigating these applications requires a model of PMD. Historically, real-world fiber links were described using weakly constrained models that did not depend on the underlying physical mechanisms of PMD \cite{OldModel1, OldModel_TimeStochastic, OG_PMD_Analysis, CzeglediOFC}. A more physically constrained model would be more useful across applications. For fiber sensors, such a model could shed light on the consequences of environmental noise such as temperature changes or vibrations. In applications where PMD mitigation is needed, modeling can tie observed PMD to the location of time-varying elements, such as a data closet with poor temperature control or a haphazardly dangling fiber length buffeted by HVAC turbulence. 
In quantum networks, a physical fiber model could predict the contribution of PMD to entanglement infidelity for different qubit encodings (e.g.  polarization, frequency or time-bin) and examine the feasibility of approaches to PMD compensation. Tying PMD to physical parameters such as temperature is critical to many of these applications. 

In this work, we construct a first-principles model of PMD in optical fibers, in which all parts of the fiber and its behavior in time are modeled by fiber geometries, materials, and stresses. The model is implemented in an open-source Python library \cite{Git} known as the Birefringence in Fiber: Research and Optical Simulation Toolkit (BIFROST). This toolkit is a bridge between prior literature and the needs of current application domains. (In this work, we will use ``BIFROST'' to refer to both the Python implementation of the model and to the model itself.) We begin by reviewing the history of optical fibers  relevant for our work. Then we describe the top-level model of BIFROST, review the physics included in the model, and enumerate the data sources used in the numerical calculation of fiber parameters. Then we compare prior work to simulations from BIFROST to validate our model. Finally, we perform a brief example investigation with BIFROST to demonstrate its utility, and discuss future directions.

\section{A Brief History of Optical Fibers}

\begin{figure}
\centering
    \includegraphics[width=0.83\columnwidth]{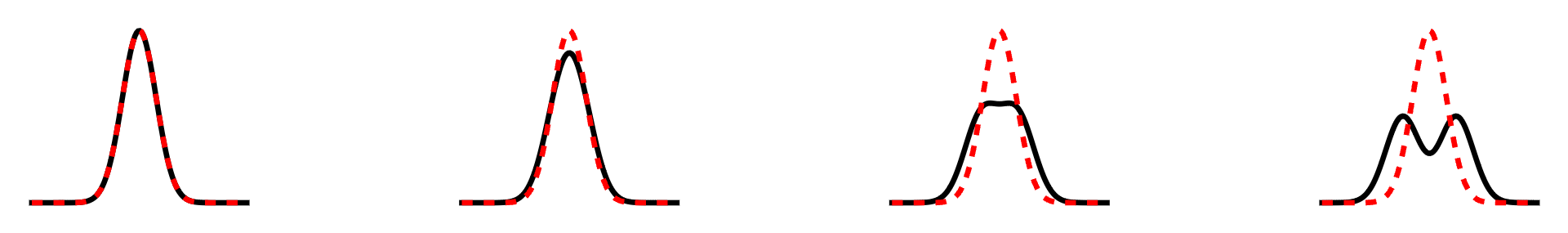}
    
    \vspace{0.5em}
    
    \includegraphics[width=\columnwidth]{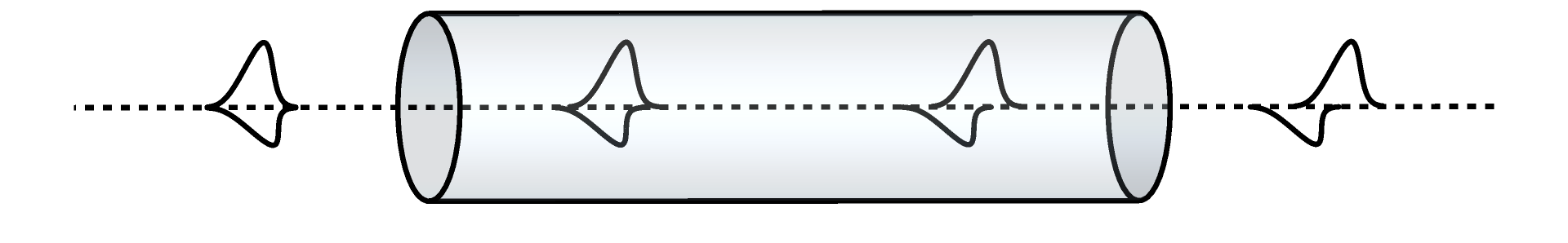}
\caption{\label{fig:DGDCartoon} Illustration of the effect of PMD on light pulses whose polarization is not aligned with the birefringence axes of a fiber. As the pulse travels through the fiber, a relative delay accumulates between different polarizations, resulting in pulse broadening. The upper panels show this broadening by comparing the output intensity profile (black solid line) to the input one (red dashed line).}
\end{figure}

Among the earliest live communication signals sent via optical fiber was telephone traffic at hundreds of Mb/s$\cdot$km, in 1977 \cite{Chraplyvy.Review}. Already by this time, birefringence in fiber was identified as a possible limit on data rates \cite{EarlyPMD1, EarlyPMD2, Chraplyvy.Review}. The reason for this is shown in Fig.~\ref{fig:DGDCartoon}: in on-off intensity schemes with  no polarization control, birefringence causes the parts of the light along the fast and slow axes to propagate at different group velocities, resulting in pulse broadening \cite{YarivYeh}. If a pulse broadens too much, it can interfere with adjacent data frames, limiting the data rate \cite{Chraplyvy.Review}. Initial measurements in short ($\approx 1$-$10$~m) fibers \cite{EarlyPMD1}, when extrapolated linearly with fiber length, suggested a PMD-limited data rate of $\sim 10$~Gb/s$\cdot$km. Subsequent direct measurements on long ($\approx 1$~km) fiber links \cite{Mochizuki} and a flurry of theoretical activity \cite[e.g.][]{EarlyPMD2, EarlyPMDTheory1, EarlyPMDTheory2} showed that PMD actually scales as the square root of the length. This increased the estimated limit due to PMD to tens of Tb/s$\cdot$km \cite{Mochizuki}.

Obtaining the capacities of modern global-scale systems (tens of Pb/s$\cdot$km for the Pacific Light Cable Network and TAT-14 \cite{Chraplyvy.Review}) required further reduction of PMD, which was obtained by fiber spinning \cite{SpinningOG, SpinningLongFiber, SpinningHistoryAndTheory, SpinningTheory2, SpinningNonlocal}. To manufacture a spun fiber, wheels in contact with the glass apply torques while the glass is soft (during the fiber draw stage), causing the fiber to rotate about its long axis; these rotations are locked in as the glass cools and hardens \cite{SpinningHistoryAndTheory}. The spin rate is typically periodic with a period of a few meters \cite{SpinningHistoryAndTheory, SpinningTheoryZeros2019}. This spinning scrambles the birefringence axes in the fiber, preventing the coherent build-up of delay between two modes. Spinning was widely deployed beginning in the late 2000s \cite{Chraplyvy.Review}. Most modern long-distance telecom fiber is spun fiber. In systems deploying spun fiber, the small remaining amount of PMD-induced broadening can be compensated by modern polarization-sensitive coherent transceivers \cite{Chraplyvy.Review, Savory_Review}. A recent revival of interest in modeling PMD stems from the need to further improve transceiver compensation algorithms \cite{CzeglediOFC}.

Modeling fiber PMD began alongside and in support of the earliest experimental developments in the fiber telecommunications industry. The original way to model PMD in long fibers, going back to the 1980s, was to break the fiber up into many sections and specify different birefringence magnitudes and orientations for each segment \cite{EarlyPMD2, OldModel1, OldModel2, OldModel_TimeStochastic, OG_PMD_Analysis}. Such models were able to reproduce certain observed statistical properties of PMD, but they suffered from two major challenges: time dependence was difficult to account for, often requiring stochastic calculus methods \cite{OldModel_TimeStochastic}, and the model was not well-motivated physically: the random distributions of birefringences and orientations did not meaningfully correspond to the physical reality of the fiber. 

\begin{figure*}
\centering
   \includegraphics[width=\textwidth]{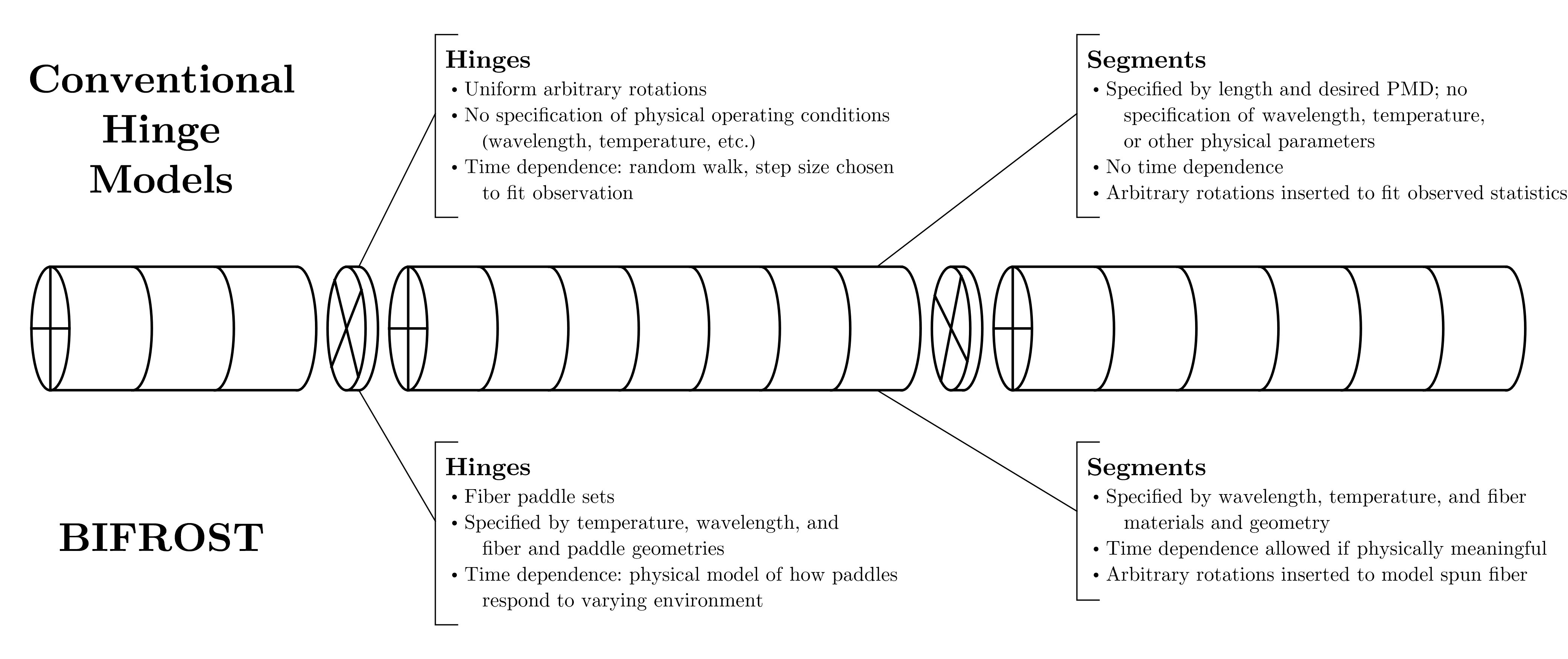}
\caption{\label{fig:HingeModelsComp} A comparison between conventional hinge models \cite{HingeOG, CzeglediSciRep, CzeglediOFC} and BIFROST. Both models rely on splitting the fiber into segments juxtaposed with hinges, but the specifications and behaviors of these elements are fundamentally different.}
\end{figure*}

An improvement on the situation, first introduced in 2006 and widely adopted since, is the so-called hinge model \cite{HingeOG}. This model, shown in Fig.~\ref{fig:HingeModelsComp}, describes buried fibers as a set of long segments alternating with short lengths called ``hinges.'' The long segments correspond to the buried parts of the fiber, and are modeled as stable in time because underground environmental variations are small. Hinges, on the other hand, correspond to above-ground segments, e.g. at repeater stations, data closets, and points of presence, which experience environmental fluctuations and resulting birefringence changes. In the hinge model, all of the time variation is in the hinges; this variation in time has been traditionally modeled either with a predetermined rotation function \cite{HingeOG} or with uniform random walks on the Poincar\'e sphere \cite{CzeglediSciRep, CzeglediOFC} with step size chosen to fit to observed statistics. In these implementations, the long time-invariant segments are specified by a fixed total length and a desired PMD, and are subdivided by arbitrary fixed rotations to reproduce observed statistical distributions of PMD. The hinge model has had some success in reproducing the observed output statistics of PMD in buried fibers \cite{HingeOG, CzeglediOFC}, but it is limited: there is no connection to physical parameters such as operating temperature or external stress, making it impossible to explore how a fiber link responds to environmental changes.

\section{The Top-Level Model of BIFROST}

In BIFROST, we adapt the hinge model to include physically motivated (and physically constrained) parameters (Fig.~\ref{fig:HingeModelsComp}). We model fiber as a step-index cylindrical waveguide, deriving refractive indices from material properties and environmental variables such as temperature. Birefringences from multiple physical mechanisms are included. Long, stable segments are specified by fiber geometry, core/cladding glass types, bend radii and twist rates where appropriate, and operating temperature. Each hinge is modeled as a set of fiber paddles, chosen because they are common experimental tools for manipulating polarization. The paddles are defined by a radius of curvature, number of fiber turns, and paddle angle. Variation in time must be modeled according to how environmental changes, like temperature fluctuations, affect the physical parameters of the paddles. Additionally, we model spinning by inserting arbitrary Jones matrix rotations in the long fiber segments, with spacing determined by typical fiber spinning periods.

Thus, rather than specifying a \emph{desired} hinge or segment PMD, BIFROST uses physical parameters to calculate an \emph{expected} PMD. Unlike previous models, both our segments and our hinges are fully specified by physical parameters. Grounding the calculations in physical birefringence mechanisms results in a model that can describe the changes in the output polarization due to temperature, wavelength variation, and more, making BIFROST a powerful predictive tool.

\section{A Review of Optical Fiber Physics}

To connect physical parameters to fiber link PMD, BIFROST requires models of the relevant physics, including the propagation of the guided mode in the fiber, chromatic dispersion effects, and the sources of birefringence. We detail these models in this section. 

\subsection{Modes of a Cylindrical Waveguide}

We model fibers as step-index cylindrical waveguides \cite{SnyderLove}, whose refractive index profile is
\begin{equation}
    n(r) = \left\{ \begin{array}{cc} 
    n_\text{co} & 0 \leq r<r_0, \\
    n_\text{cl} & r_0<r<r_\text{cl}
    \end{array} \right.,
\end{equation}
where $r_0$ [m] is the core radius, $r_\text{c}$ [m] is the cladding radius, and $n_\text{co} > n_\text{cl}$ is required to obtain guided modes.

In the weakly guiding limit $n_\text{co} - n_\text{cl} \ll 1$, the modes in an optical fiber are nearly transverse modes \cite{YarivYeh}. For the fundamental mode, with some approximation \cite{Gloge}, the propagation constant $\beta$ [1/m] can be written
\begin{equation}
    \beta^2 = n_\text{co}^2k_0^2 - \frac{1}{r_0^2} \left[ \frac{(1+\sqrt{2})V}{1 + (4 + V^4)^{1/4}} \right]^2,
    \label{eqn:BetaFiberFinal}
\end{equation}
where $k_0=2\pi/\lambda$ [1/m] is the free-space wavenumber and
\begin{equation}
     V \equiv r_0 k_0 (n_\text{co}^2-n_\text{cl}^2)^{1/2}
     \label{eqn:VFiber}
\end{equation}
is the normalized (dimensionless) frequency parameter. The fiber is single-mode when $V < 2.405$ (of course, this means only one \emph{spatial} mode propagates in the fiber; there are \emph{two} orthogonally polarized modes).

In optical fibers, a small core-cladding refractive index difference enables efficient total internal reflection and reduces losses, so virtually all commercial fibers are in the weakly guiding limit, with refractive index differences commonly less than 1\%. Though many deployed fibers have a graded index profile, we only simulate step-index fiber in this work.

\subsection{Chromatic Dispersion}

Chromatic dispersion (CD), the variation of the propagation constant with wavelength, comes from two sources \cite{BuckOpticalFibers}. One source is waveguide CD, which occurs because the fraction of the guided mode in the core varies with wavelength; Eqn.~\ref{eqn:BetaFiberFinal} shows the resulting variation of $\beta$ with wavelength. The other source is material CD, which occurs when $n_\text{co}$ and $n_\text{cl}$ vary with wavelength, as happens in all real materials.

\begin{figure}
    \centering
    \includegraphics[width=\columnwidth]{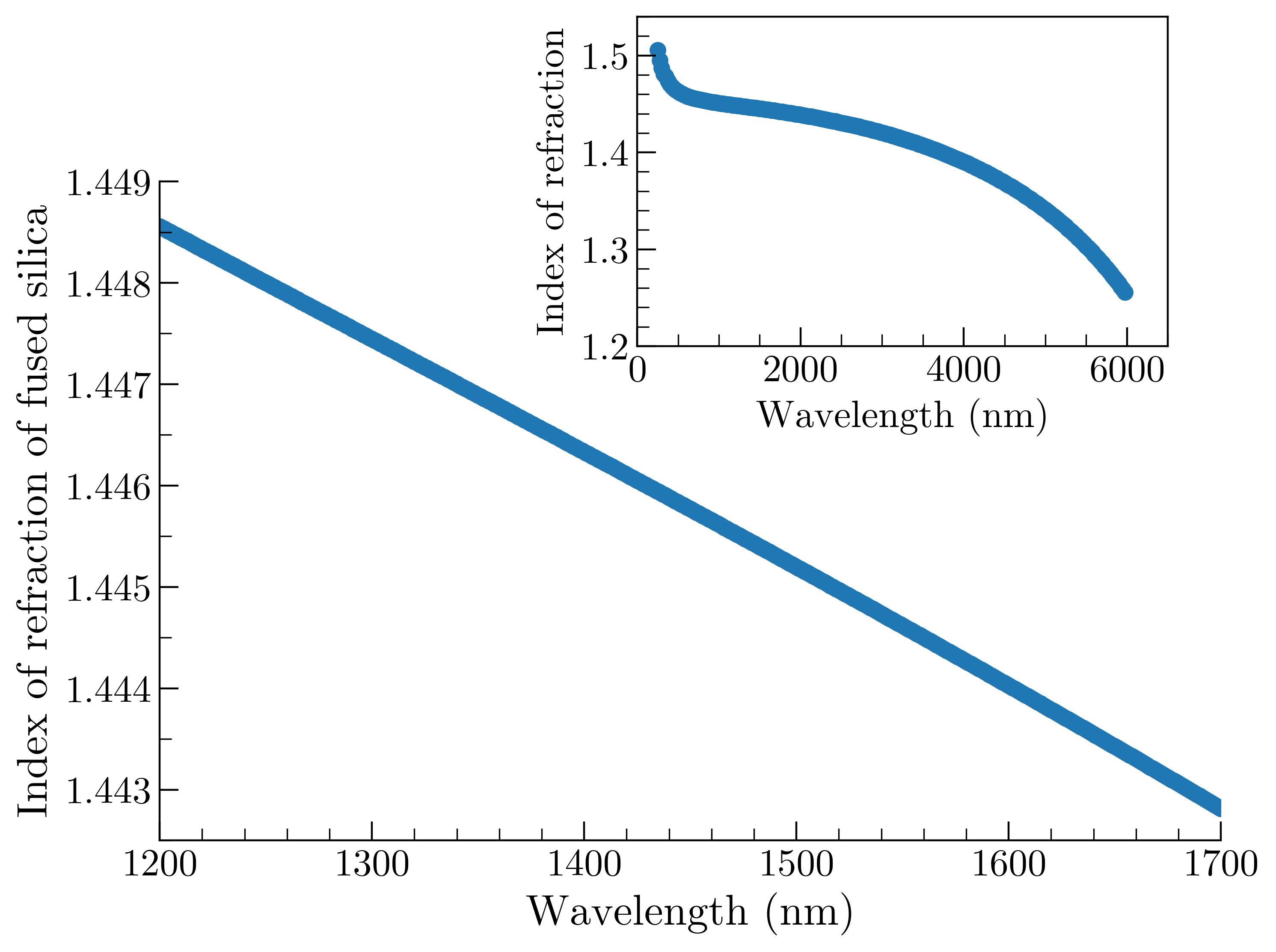} 
    
    \includegraphics[width=\columnwidth]{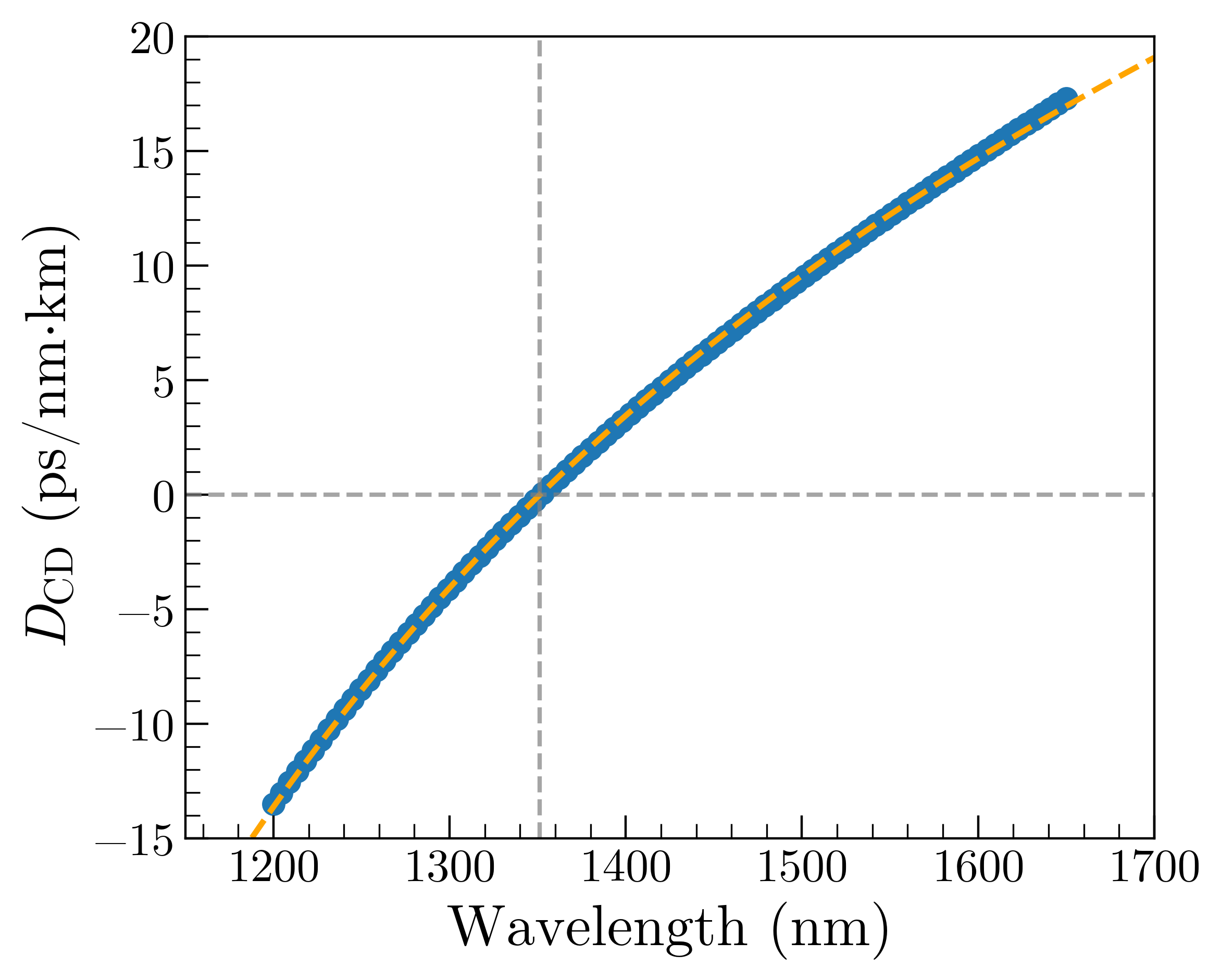} 
    
    \caption{\label{fig:SilicaCD} Chromatic dispersion of fused silica. Top: the index of refraction of fused silica, using the measurements of Ref.~\cite{FusedSilica_ThreePole_Temp2} at 20${}^\circ$C. Bottom: calculated $D_\text{CD}$ (thick blue line) using Eqns.~\ref{eqn:BetaFiberFinal} and \ref{eqn:Sellmeier} with the data from the top panel. The orange dashed line indicates the best fit to the heuristic Eqn.~\ref{eqn:Heuristic_D_CD}, showing a visually good fit to the calculated values. The gray dashed lines indicate the zero-dispersion wavelength. For this figure, we used a core radius $4.1\,\mu$m, a temperature of $20^\circ$C, a pure fused silica core, and a cladding doped with 3.6\% germania. (We discuss the numerical implementation of the doping in later sections.)}
\end{figure}

A common model of material CD is the Sellmeier model, in which the material is treated as having absorptive resonances which are approximated as delta functions in the imaginary part of the material's electric susceptibility. The Kramers-Kr\"onig relations tie these resonances to the real part of the susceptibility and thus the refractive index, which can be written as \cite{BuckOpticalFibers}
\begin{equation}
    n^2(\lambda) = 1 + \sum_i \frac{B_i \lambda^2}{\lambda^2 - C_i^2},
    \label{eqn:Sellmeier}
\end{equation}
where $B_i$ [dimensionless] are resonance strengths and $C_i$ [m] are the resonance wavelengths, and the sum runs over as many resonances as are practically measurable or necessary for the intended application. In fused silica, usually either three \cite{FusedSilica_Sellmeier_ThreePole, FusedSilica_ThreePole_Temp1, FusedSilica_ThreePole_Temp2} or two \cite{FusedSilica_TwoTerm_Temp, FusedSilica_TempDependence1} resonances are used. In both cases, at least one resonance in the deep UV captures the effect of photons promoting electrons to higher energy bands of the material, while at least one resonance in the infrared captures excitation of vibrational modes of the glass \cite{FusedSilica_TwoTerm_Temp}. The refractive index of fused silica is shown in the top panel of Fig.~\ref{fig:SilicaCD}, indicating both a UV resonance at the short wavelengths and the decrease in $n$ due to lattice absorption at long wavelengths.

The total chromatic dispersion of an optical fiber is often specified by the group velocity dispersion parameter $D_{\text{CD}}$ [s/m${}^2$]. There is a heuristic expression for $D_{\text{CD}}$ that combines both waveguide and material dispersion that works well for fused silica fibers \cite{FusedSilica_TempDependence1}:
\begin{equation}
    D_{\text{CD}}(\lambda) = \frac{S_0}{4} \left( \lambda - \frac{\lambda_0^4}{\lambda^3} \right)
    \label{eqn:Heuristic_D_CD}
\end{equation}
where $\lambda_0$ [m] is the zero-dispersion wavelength and $S_0$ [s/m${}^3$] is the slope of $D_\text{CD}$ at $\lambda_0$. It is common to find these two parameters in fiber datasheets. The bottom panel of Fig.~\ref{fig:SilicaCD} shows $D_{\text{CD}}$ calculated using Eqns.~\ref{eqn:BetaFiberFinal} and \ref{eqn:Sellmeier} (thick blue line) compared with the heuristic Eqn.~\ref{eqn:Heuristic_D_CD} (orange dashed line). The heuristic expression captures the behavior well.

Because there is both material and waveguide dispersion, it is possible to design the fiber so that the two effects cancel at a desired wavelength, for instance at 1550 nm rather than the typical $\sim 1310$~nm. This so-called dispersion-compensating fiber can be developed by changing the materials used in the core and cladding and by engineering the refractive index profile \cite{DCF1, DCF2}. Though it may be a useful future direction, we do not model dispersion-compensating fiber in the present work.

\subsection{Sources of Birefringence}

To model birefringences, we choose coordinates such that the fiber lies along the $z$-axis, and the eigenmodes have linear polarizations along the $x$- and $y$-axes. These modes have indices of refraction $n_x$ and $n_y$ with difference $\Delta n = n_y - n_x$. In the weakly guiding limit, it is acceptable to approximate the resulting propagation constant difference as $\Delta \beta \equiv \beta_y - \beta_x = k_0 \Delta n$ [1/m], where $k_0$ is the free-space wavenumber.

While we can use the birefringence magnitudes directly to compare different fibers or different birefringence mechanisms, it is also useful to define a polarization beat length $L_\text{p} = 2\pi/\Delta\beta$ [m], which is the characteristic length at which the phase difference between two reference polarizations becomes $2\pi$. This beat length can be useful when thinking about other characteristic lengths of a given application, such as the spin period of a spun fiber or the length scale of a disturbance in a sensing application.

In this work, we model four physical sources of birefringence: core ellipticity, asymmetric thermal stress, bending, and twisting.

\paragraph{Core Ellipticity.} When the core of the fiber is not circular, the boundary conditions that determine the propagation constants are different. Letting the ellipse be oriented with its semimajor and semiminor axes $r_{x,y}$ along the $x$- and $y$-axes (explicitly, $r_x > r_y$), we obtain two effective $V$ parameters $V_{x,y} = k_0r_{x,y} \sqrt{n_\text{co}^2 - n_\text{cl}^2}$. We also define the eccentricity $e^2 = 1-r_y^2/r_x^2$ \footnote{If $\rho_x < \rho_y$, all of the expressions in this section flip $e^2 \rightarrow -e^2$.}. For small core eccentricities, we can use the Gaussian approximation to derive the resulting birefringence \cite{SnyderLove}. For $V>1$ and to second order in $e$, it is \cite{SnyderLove, Mabrouki_FiberModeling, Boudrioua_FiberModeling}
\begin{equation}
    \Delta\beta = -\frac{e^2 (2\Delta)^{3/2}}{\rho} \frac{4}{V^3} \frac{(\ln V)^3}{1 + \ln V},
    \label{eqn:B_CNC}
\end{equation}
where $\rho=\sqrt{r_x r_y}$ and $V = \sqrt{V_xV_y}$ are the core radius and normalized frequency parameter for the circular fiber of area equal to our elliptical one, and
\begin{equation}
    \Delta \equiv \frac{1}{2} \left( 1 - \frac{n_\text{cl}^2}{n_\text{co}^2} \right) \approx \frac{n_\text{co} - n_\text{cl}}{n_\text{co}},
\end{equation}
where the approximation is good for weakly guiding fibers.

\paragraph{Asymmetric Thermal Stress.} The core and cladding, being different materials, may have different coefficients of thermal expansion $\alpha_\text{co}$ and $\alpha_\text{cl}$ [1/K]. As a result, temperature changes place a stress on the fiber core along the length of the fiber; when the core is also noncircular, this thermal stress becomes asymmetric, resulting in a birefringence. The stress is calculated relative to the softening temperature $T_\text{S}$ [K], above which the glass is soft and able to expand or compress without adding stress, and below which the glass is hard and resistant to compression from cooling. The asymmetric thermal stress results in a birefringence \cite{ThermalStressBirefringence, OG_PMD_Analysis}
\begin{align}
    \label{eqn:B_ATS}
    \Delta\beta &= -k_0\left( 1 - \frac{u^2}{V^2} \right) \\
    &\hspace{0.75em}\times\left[ \frac{1}{2} n_\text{co}^3 (p_{11} - p_{12}) \frac{(\alpha_\text{cl}-\alpha_\text{co}) (T_\text{S} - T)}{1-\nu_\text{p}^2} \frac{r_x - r_y}{r_x + r_y} \right]. \nonumber
\end{align}
Here $T$ is the operating temperature of the fiber, $u=r_0 (n_\text{co}^2 k_0^2 - \beta^2)^{1/2}$, $p_{11}$ and $p_{12}$ are the dimensionless photoelastic constants of the core, $\alpha_\text{co}$ and $\alpha_\text{cl}$ are the thermal expansion coefficients for the core and cladding, and $\nu_\text{p}$ is the dimensionless Poisson's ratio for the core.

\paragraph{Bending.} When a fiber is bent, there is a stress toward the center of the bend which causes refractive index changes. If the bend is in the $xz$ plane and its radius of curvature $R$ [m] is much greater than the radius of the cladding $r_\text{cl}$, then the resulting birefringence can be written
\begin{equation}
    \Delta\beta = k_0 \frac{n_\text{co}^3}{4}(p_{12}-p_{11})(1+\nu_\text{p}) \frac{r_\text{cl}^2}{R^2} = \frac{1}{2} C_\text{s} \frac{r_\text{cl}^2}{R^2},
\end{equation}
where $C_\text{s}/k_0 = (n_\text{co}^3/2) (p_{11} - p_{12}) (1+\nu_\text{p})$ [dimensionless] is a commonly defined combination of material constants loosely called the ``strain-optic coefficient.'' 

If the bending involves some axial tension, such as tight wrapping around a drum, then there is an additional birefringence \cite{TensionBendingBirefringence}
\begin{equation}
    \Delta\beta = C_\text{s} \frac{2 - 3\nu_\text{p}}{1 - \nu_\text{p}} \frac{r}{R} \frac{F}{\pi r^2 E},
    \label{eqn:B_BND}
\end{equation}
where $E$ [Pa] is the Young's modulus of the fiber and $F$ [N] is the tension force.

\paragraph{Twisting.} When a fiber is twisted, a shear stress is introduced, which couples the transverse field of one mode to the longitudinal field of another. (Because we work in the weakly guiding limit, we have largely been ignoring the longitudinal field components, but in a typical Corning SMF-28e fiber at 1550 nm, the longitudinal field amplitude is a few percent of the transverse field amplitude.) The transverse and longitudinal fields are $\pi/2$ out of phase, so the coupling results in a circular birefringence. The birefringence is proportional to the twist rate $\tau$ [rad/m], and can be written \cite{Rashleigh}
\begin{equation}
    \beta_{\text{LC}} - \beta_{\text{RC}} = -\frac{1}{2} n_\text{co}^2 (p_{11} - p_{12}) \tau
    \label{eqn:B_TWS}
\end{equation}
where LC and RC stand for left and right circular modes, respectively.

\paragraph{Additional mechanisms.} While we do not model other mechanisms in this work, we note that ellipticity of the cladding, non-concentric cladding and core, external asymmetric stress, transverse electric fields, and axial magnetic fields \cite{Rashleigh} are all additional possible contributors to fiber birefringence. The inclusion of these mechanisms in BIFROST is a direction for future work.

\subsection{Polarization Transfer Functions with Jones Matrices}

To express polarization mathematically, we use the Jones formalism \cite{YarivYeh}. Given a transverse light wave, the field components in the $x$- and $y$-directions can be written
\begin{subequations}
\begin{align}
    E_x &= A_x \cos(\beta_0 z - \omega t + \delta_x), \\
    E_y &= A_y \cos(\beta_0 z - \omega t + \delta_y).
\end{align}
\label{eqn:BaseWaveNotations}
\end{subequations}
Here $\beta_0$ is a common propagation constant. For birefringent elements, we put the birefringence into the phases $\delta_{x,y}$, which take on position-dependence (e.g. for linear birefringence aligned with the $x$- and $y$-axes, we have $\delta_{x,y} = \beta_{x,y} z$). The Jones vector for this light wave is defined as
\begin{align}
    \vec{V} = \left[ \begin{array}{c} A_x \\ A_y e^{i\delta} \end{array} \right]
\end{align}
with $\delta = \delta_y - \delta_x$.

Jones vectors express states of polarization (SOPs). The transformation of the SOP as light propagates through a medium is described by the Jones matrix of the medium: $\vec{V}_{\text{out}} = J\vec{V}_{\text{in}}$. When a medium is composed of multiple elements each having constant birefringence along its length, with Jones matrices $J_i$, then the total Jones matrix is the product of the individual matrices,
\begin{equation}
    J_{\text{total}} = \prod_i J_i.
    \label{eqn:Jtot}
\end{equation}
The order of multiplication is important and should correspond to the order in which the light encounters the segments. In our case, these elements include the long stable fiber segments, the segments of fiber that make up the fiber-paddle-modeled hinges, and the rotators used to model spun fiber. A medium's total Jones matrix is an expression of its polarization transfer function (PTF).

Fiber PTFs can be measured in several ways \cite{PMD_PNASReview}. A common method is the Mueller matrix method \cite{MMM_OG}, in which the output SOP is measured for two linearly independent SOPs, from which the full PTF can be reconstructed.

For a linearly birefringent medium of length $L$, which adds only a different phase to the $x$- and $y$-components of the light, the Jones matrix can be written \cite{YarivYeh, OG_PMD_Analysis}
\begin{equation}
    J_\text{B,lin} = \left(
    \begin{array}{cc}
    e^{i\Delta\beta \,L/2} & 0 \\
    0 & e^{-i\Delta\beta\,L/2}
    \end{array}
    \right).
    \label{eqn:JBLin}
\end{equation}
Here we ignore the common phase shift imparted on both modes.

A fiber element that has circular birefringence, such as occurs in twisted fibers \cite{CircularBiref, CircularBiref2}, can be written \cite{YarivYeh}
\begin{equation}
    J_\text{B,circ} = \left(
    \begin{array}{cc}
    \cos(\Delta\beta L/2) & -\sin(\Delta\beta L/2) \\
    \sin(\Delta\beta L/2) & \cos(\Delta\beta L/2)
    \end{array}
    \right).
    \label{eqn:JBCirc}
\end{equation}
where $\Delta\beta = \beta_{\text{LC}} - \beta_{\text{RC}}$.

\subsection{Modeling Spun Fiber}

As previously discussed, a common technique for decreasing PMD in fibers is to spin the fiber during manufacture, before the glass is cooled and solidified. The spinning is often done with a periodic spin rate $r=A\sin(\omega t)$; spinning back and forth avoids the build-up of torsional effects over the fiber. Controlling the period and amplitude of the spin rate provides control over the final fiber PMD \cite{SpinningTheory2}. It has been suggested that anisotropies such as core noncircularity introduced during manufacture are not reduced by spinning \cite{SpinningNonlocal}; that is, the way spinning decreases PMD is by scrambling axes, not by reducing birefringence magnitudes.

Here we model spun fiber by inserting arbitrary Jones matrix rotations in long segments of fiber, with spacing between the rotations that is of the order of the fiber spin period. We discuss the validity and consequences of this model in later sections; here we simply describe the implementation. To sample rotations uniformly, it is insufficient to uniformly sample rotation angles and axes; this overweights rotations with angles near $0$ and $\pi$. Instead we draw on a mathematical insight: a uniform sampling of Jones matrices is equivalent to uniformly sampling the surface of a 3-sphere (a surface that lives in $\mathbb{R}^4$). To do this \cite{MullerSampling}, we draw four numbers $g_1, g_2, g_3, g_4$ from the standard normal distribution and form the vector $\vec{g}=(g_1, g_2, g_3, g_4)$. Then we rewrite the normalized vector as $\vec{g}/\|\vec{g}\| = (\cos\theta, a_1\sin\theta, a_2\sin\theta, a_3\sin\theta)$, such that $\theta$ is a rotation angle and $(a_1, a_2, a_3)$ is a rotation axis. The Jones matrix corresponding to this sample is 
\begin{equation}
   J(\theta, \vec{a}) = I_2 \cos\theta - i \vec{a} \cdot \vec{\sigma} \sin\theta
   \label{eqn:JRotator}
\end{equation}
where $\vec\sigma = (\sigma_x,\sigma_y, \sigma_z)$ is the Pauli vector and $I_2$ is the $2\times2$ identity matrix. The family of Jones matrices sampled in this way implements uniform rotations on the Poincar\'e sphere \cite{CzeglediSciRep, CzeglediOFC}.

\subsection{Differential Group Delay}

While the Jones matrix of the fiber link supplies its entire PTF at one wavelength, the telecommunications industry has typically only been interested in PMD-induced pulse broadening, which is a single number, known as the differential group delay (DGD) [s] \cite{YarivYeh, PMD_PNASReview}. It is common (including in many fiber specification sheets) to specify PMD by the DGD; it is a quantification of the phenomenon shown earlier in Fig.~\ref{fig:DGDCartoon}.

For a length $L$ of fiber whose birefringence is constant along its length, the DGD is \cite{YarivYeh}:
\begin{equation}
    \tau_{\text{DGD}} = \frac{L}{v_\text{g,s}} - \frac{L}{v_\text{g,f}}
\end{equation}
where $v_\text{g,s}$ and $v_\text{g,f}$ are the slow and fast mode group velocities [m/s]. Using $1/v_\text{g} = dk/d\omega$, $k=n(\omega)\omega/c$, and $\Delta\beta = k_0\Delta n$, we can write the DGD as
\begin{equation}
    \tau_{\text{DGD}} = L \frac{d\,\Delta\beta}{d\omega}.
    \label{eqn:DGDOmegaDerivative}
\end{equation}
This expression implies that, while DGD is intuitive in a time-domain picture, there is an intimately connected frequency-domain picture. 

To support this point, let us consider what happens when we keep the input polarization constant but vary the frequency of the light \cite{YarivYeh, PMD_PNASReview}. For a fiber Jones matrix $J$, the input and output SOPs are related by $\vec{V}_{\text{out}} = J\vec{V}_{\text{in}}$. If the frequency of the light is varied by a small amount, then the change in the output state can be described by $\partial \vec{V}_{\text{out}}/\partial\omega = (\partial J/\partial\omega) \vec{V}_\text{in}$. The matrix $\partial J/\partial\omega$, being an element of SU(2), is a rotation; the axis of the rotation defines two SOPs which do not change to first order with the frequency change. These special states, whose output Jones vectors are $\vec{V}_{\text{out}}$, both satisfy the relation
\begin{equation}
    \frac{\partial \vec{V}_{\text{out}}}{\partial\omega} = i\delta \vec{V}_{\text{out}}
    \label{eqn:DGDeigenvalue}
\end{equation}
for some constant $\delta$ (which has units of time and is real when $J$ is unitary), representing a phase gained through the fiber (which does not change the SOP). The output state is simply $\vec{V}_{\text{out}} = J\vec{V}_{\text{in}}$ for a fiber Jones matrix $J$, so we find
\begin{equation}
    J^{-1} \frac{\partial J}{\partial\omega} \vec{V}_{\text{in}} = i\delta \vec{V}_{\text{in}}
    \label{eqn:DGDFreqEigenequation}
\end{equation}
which is an eigenvalue problem for the matrix $J^{-1} (\partial J/\partial\omega)$. In the case where the fiber has constant linear birefringence, Eqn.~\ref{eqn:JBLin} yields two eigenvalues $\pm\delta_0$, where
\begin{equation}
    \delta_0 = \frac{L}{2}\left( \frac{\Delta n}{c} + \frac{\omega}{c} \frac{d\Delta n}{d\omega} \right) = \frac{L}{2} \frac{d\,\Delta\beta}{d\omega},
\end{equation}
such that the total delay between these two modes is exactly $2\delta_0 = \tau_{\text{DGD}}$ from Eqn.~\ref{eqn:DGDOmegaDerivative}. The eigenmodes in this case are $(\,1\,\,\,\,0\,)^T$ and $(\,0\,\,\,\,1\,)^T$ (where ${}^T$ is the transpose), which are sensible as these states are aligned with the fast and slow axes of the fiber. The eigenmodes are called principal states of polarization; they do not vary to first order with frequency, and the time delay between them is exactly the DGD of the time-domain picture.

What is useful about this mapping between time and frequency domains is that the time-domain picture is only simple when the birefringence is constant; when there are many elements with their own birefringences, keeping track of time delays becomes cumbersome. The frequency-domain picture we have described above relies only on the total Jones matrix; the DGD can always be calculated from Eqn.~\ref{eqn:DGDFreqEigenequation}. Thus, for real fibers, and for many PMD measurement methods, the frequency-domain picture is the  sensible one \cite{PMD_PNASReview}. 

The DGD is a single scalar value, and therefore insufficient to describe the full PTF. Nevertheless, the DGD is still helpful as a heuristic metric of polarization rotation in the fiber. To see this, let us consider putting a temporally long pulse through each of three different fibers and consider the consequence of changing the fiber temperature.  (1) If the fiber has no birefringence, then there is no PMD, and therefore zero DGD, and also the SOP remains the same at the output as at the input. The temperature change has no effect on the output SOP. (2) Conversely, in a long segment of fiber with constant birefringence, the delay between the two polarization components is allowed to build up over a long distance, so the total DGD is large. At the same time the SOP is also rotating along the fiber length, so the output SOP is very sensitive to changes in temperature. (3) Finally, for a long spun fiber, the spinning scrambles the birefringence axes; thus no two polarization modes are allowed to build up large relative delays, and the total DGD is small (in comparison to an unspun fiber of the same length). As the light traverses the fiber, the SOP does a random walk on the Poincar\'e sphere, rotating around one axis for a short length, then another axis for a short length, and so on. Such a random walk is hard to track, but the polarization does not wander as quickly as a function of length down the fiber. When the temperature is changed, each of these random walk steps is lengthened or shortened slightly, making it challenging to predict how the output SOP changes, but since the changes of length are in arbitrary directions, the total change to the SOP is smaller than for the unspun fiber of the same length.

This discussion suggests two things. First, the DGD can be an indicator of how fast the SOP changes as a function of environmental conditions such as temperature. Second, cases (2) and (3) indicate that, although the DGD is lower, the SOP reacts far less predictably to a change of environmental conditions. Thus, fiber spinning, while being an excellent solution to the problem of classical pulse broadening, makes the situation about as bad as possible for tracking the SOP. The latter task is required for instance in polarization-encoded quantum networks.

\section{Calculation Data and Methods}

Our aim in this work is to model typical commercial telecommunications fiber links. Much of this fiber is predominantly silica-germania glass \cite{TempDependenceSiGe}, i.e. pure silica (SiO${}_2$) for the cladding, and silica doped with a small amount of germania (GeO${}_2$) for the core, so we model these materials in our work. (Future extensions to BIFROST may include adding additional materials.) The material properties of bulk silica and bulk germania have been well-studied, giving us a wealth of data to use in BIFROST.  To specify the effects of germania doping, we use the mole percent (\%mol) (the fraction of molecules in the glass that are germania \cite{TempDependenceSiGe}) to specify the concentration, and we use a linear additive model for each of the relevant fiber properties. That is, for a given fiber property $Y$ which has a value $Y_\text{s}$ for pure silica and $Y_\text{g}$ for pure germania, the value we use for silica with molar fraction $m$ of germania doping is
\begin{equation}
    Y = (1-m)Y_\text{s} + mY_\text{g}.
    \label{eqn:Additive}
\end{equation}
This kind of linear additive model has been used for nearly all of the properties we need, including the refractive index \cite{TempDependenceSiGe}, thermal expansion coefficient \cite{CTEVals, OG_PMD_Analysis}, softening temperature \cite{OG_PMD_Analysis, MiscFiberProps1}, Poisson's ratio \cite{OG_PMD_Analysis, MiscFiberProps1}, and Young's modulus \cite{TempDependenceSiGe}. (We also adopt such a model for the photoelastic constants.) This model is good when the doping concentration is small; at larger concentrations of germania, there are some corrections to a linear model, e.g. a quadratic term in the refractive index \cite{TempDependenceSiGe}. The values we use for pure silica and pure germania are given in Table \ref{tab:FiberProps}.

\begin{table}
\caption{\label{tab:FiberProps} Material properties used in BIFROST, for pure silica and pure germania, with references. Listed are the coefficient of thermal expansion (CTE), the softening temperature $T_\text{S}$, Poisson's ratio $\nu_\text{p}$, Young's modulus $E$, and the photoelastic constants $p_{11}$ and $p_{12}$.}
\renewcommand{\arraystretch}{1.2}
\begin{ruledtabular}
\begin{tabular}{lccr}
Quantity & Silica & Germania & Refs. \\
\colrule
CTE (K${}^{-1}$)\footnote{The CTE has a temperature dependence which we neglect \cite{TempDependenceSiGe}.} & $5.4 \times 10^{-7}$ & $10 \times 10^{-6}$ & \cite{CTEVals, OG_PMD_Analysis} \\
$T_\text{S}$ (${}^\circ$C) & $1100$ & $300$ & \cite{OG_PMD_Analysis} \\
$\nu_\text{p}$ & $0.170$ & $0.212$ & \cite{TempDependenceSiGe, MiscFiberProps2} \\
$E$ (GPa) & $74$ & $45.5$ & \cite{MiscFiberProps1, MiscFiberProps2} \\
$p_{11}$\footnote{This coefficient has some wavelength and temperature dependence which we neglect \cite{MiscFiberProps1, StressOptic_LambdaTempDependence, TempDependenceSiGe}.} & $0.121$ & $0.130$ & \cite{TempDependenceSiGe} \\
$p_{12}$\footnotemark[2] & $0.270$ & $0.288$ & \cite{TempDependenceSiGe}
\end{tabular}
\end{ruledtabular}
\end{table}

To calculate indices of refraction, we use three-term Sellmeier equations for both silica and germania. For silica, we use the Sellmeier coefficients from Ref.~\cite{FusedSilica_ThreePole_Temp2}. The authors measured the refractive index of Corning 7980 fused silica in the ranges $30\text{~K} \leq T \leq 300\text{~K}$ and $0.4~\mu\text{m} \leq \lambda \leq 2.6~\mu\text{m}$, and fit the data to a set of three Sellmeier resonances with quartic temperature dependence. Thus the Sellmeier equation for silica takes into account the temperature dependence of the refractive index. For germania, we use the data from Refs.~\cite{GeNVals, GeNVals2}, measured at an average of 24${}^\circ$C over the wavelength range from 365~nm to 4.3~$\mu$m. The authors do not know of any literature on the temperature-dependence of the coefficients of a three-term Sellmeier model of germania (though see Ref.~\cite{TwoTermGermaniaTempDependence} for the two-term case), but Ref.~\cite{TempDependenceSiGe} gives an expression, fitted from measurements, for the thermo-optic coefficient $dn/dT$ of bulk germania glass in the range 200-300~K at 1550~nm. We therefore calculate the refractive index of germania at a temperature $T$ by calculating the expected value at 24${}^\circ$C from Refs.~\cite{GeNVals, GeNVals2} and adding the expected change in the refractive index $\int_{24^\circ}^{T} dn(T')/dT'\,dT'$ using the thermo-optic coefficient expression of Ref.~\cite{TempDependenceSiGe}.

These data, combined with the birefringence models described previously, are sufficient to calculate the total Jones matrix of a long optical fiber. The total Jones matrix is the product of the Jones matrices for each element, Eqn.~\ref{eqn:Jtot}. Each element is either an arbitrary rotation, implemented as described above, or a birefringent element, whose birefringence is calculated from Eqns.~\ref{eqn:B_CNC}, \ref{eqn:B_ATS}, \ref{eqn:B_BND}, and \ref{eqn:B_TWS}. These formulas require both the user inputs (such fiber geometry and operating environment) and the material data described above.

Once the total Jones matrix is obtained, we calculate the DGD numerically by discretizing the derivative in Eqn.~\ref{eqn:DGDeigenvalue}. We write $\partial J/\partial \omega \approx [J(\omega + d\omega) - J(\omega)]/d\omega$ for small $d\omega$. Then Eqn.~\ref{eqn:DGDFreqEigenequation} becomes
\begin{align*}
    [J^{-1}(\omega) J(\omega + d\omega) - (1+i\delta \,d\omega)] \vec{V}_{\text{in}} = 0.
\end{align*}
That is, the eigenvalues of the matrix $J^{-1}(\omega) J(\omega + d\omega)$ are $1 + i\delta_{1,2}\,d\omega \equiv \rho_{1,2}$. The DGD is the difference of these eigenvalues, $\tau_{\text{DGD}} = \delta_1 - \delta_2$. To find this difference, we note that, if $\delta_{1,2}\,d\omega \ll 1$, we can approximate $1 + i\delta\,d\omega \approx e^{i\delta\,d\omega}$. Then 
\begin{equation}
    \tau_{\text{DGD}} = \left| \frac{\arg(\rho_1/\rho_2)}{d\omega} \right|.
    \label{eqn:CalcDGD}
\end{equation}
We use a small enough $d\omega$ for the estimate to converge (typically 0.01 nm). Note that the requirement $\delta_{1,2}\,d\omega \ll 1$ can be relaxed to $\tau_\text{DGD} \,d\omega \ll 1$ because none of the measurements involved is sensitive to absolute optical phase \cite{Heffner}.

\section{Validation of BIFROST}

Having assembled a model of PMD in optical fibers, we need to verify that this model reproduces experimentally observed results. In some sense there is a wealth of data against which we can compare BIFROST results, since many works have made measurements of PMD in optical fibers (e.g. Refs.~\cite{PMDMeas1, PMDMeas2, PMDMeas3, HingeOG} from the telecommunications engineering literature and Refs.~\cite{PMDMeas_Quantum1, PMDMeas2_Quantum, QunnectEntanglement} from the more recent quantum network literature). But in a more rigorous sense, the necessary experimental data for validation is scarce. BIFROST brings together many different ``submodels,'' including the model of the propagation constant of the fundamental modes Eqn.~\ref{eqn:BetaFiberFinal}, the models of different birefringence mechanisms Eqns.~\ref{eqn:B_CNC}, \ref{eqn:B_ATS}, \ref{eqn:B_BND}, and \ref{eqn:B_TWS}, and the linear additive model for germania doping Eqn.~\ref{eqn:Additive} described in the last section. Ideally, each of these submodels should be compared to experiments which separate out the effects of the submodel under test, which is challenging experimentally. Such experiments, to the extent it is even possible to perform them, are rare in the literature. For this reason, validation of BIFROST is ongoing, and we are explicit about what we have examined thus far. We divide validation methodologies into five categories. 

\begin{figure}
\centering
\includegraphics[width=\columnwidth]{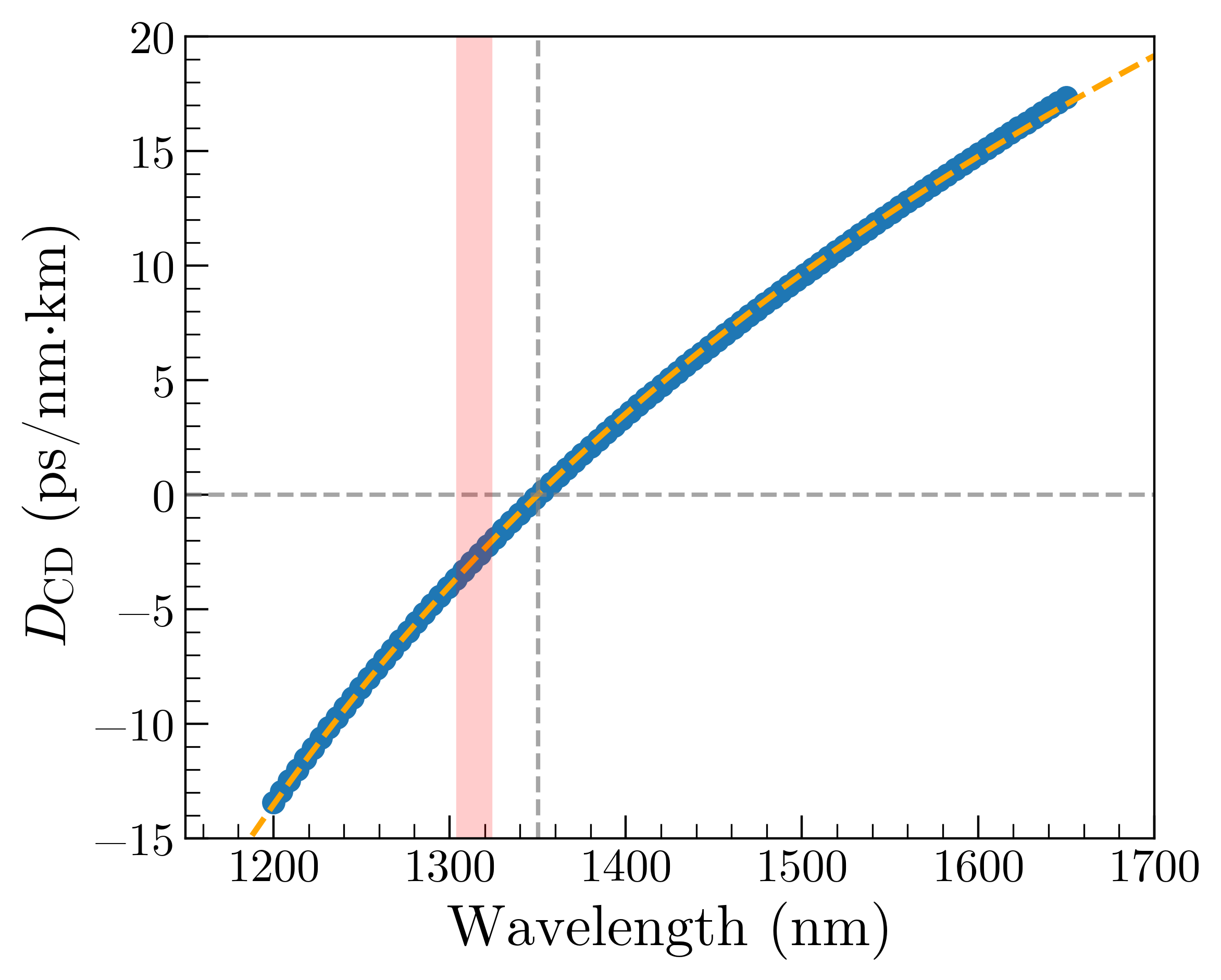}
\caption[Dispersion curve of a BIFROST-simulated fiber]{\label{fig:CorningAttemptDCD} The group velocity dispersion $D_{\text{CD}}$ of the simulated Corning SMF-28e+ fiber. The blue curve shows calculated results, and the orange dashed line is a fit to the data using the heuristic Eqn.~\ref{eqn:Heuristic_D_CD}. The red vertical band indicates the Corning specification for the zero-dispersion wavelength. The good fit to this heuristic indicates that BIFROST produces realistic behavior, but the fitted zero-dispersion wavelength is not an especially good match for the Corning SMF-28e+ specifications, as indicated in Table \ref{tab:CorningAttemptTable}.}
\end{figure}

\paragraph{Validation of intrinsic fiber properties.} One type of validation of BIFROST is the verification that the data and models of fiber material properties (i.e. Table~\ref{tab:FiberProps} and Eqn.~\ref{eqn:Additive}), combined with equations for mode conditions (Eqn.~\ref{eqn:BetaFiberFinal}), result in correct intrinsic fiber behaviors. As an example of this kind of validation, we attempt to model an exemplar commercial telecom fiber, Corning SMF-28e+. To simulate this fiber, we follow the specification sheet \cite{CorningSMF28} by inputting core and cladding diameters of $8.2~\mu$m and $125~\mu$m respectively. The specifications also give the core-cladding refractive index difference as 0.36\%. The temperature at which this difference was measured is not given, but other measurements are specified at a reference temperature of 23${}^\circ$C, so we use that temperature. If we assume the cladding to be pure silica, then BIFROST has only one degree of freedom to match this specification, namely the molar fraction of germania doping of the core; using 3.679\% doping yields $\Delta=0.36\%$ at 1550~nm and an operating temperature of 23${}^\circ$. Given these inputs, we let BIFROST calculate various additional fiber properties given in the spec sheet, and compare. The results are listed in Table \ref{tab:CorningAttemptTable}. The group velocity dispersion $D_{\text{CD}}$ is also shown for this simulated fiber in Fig. \ref{fig:CorningAttemptDCD}.

\begin{table}
\caption{\label{tab:CorningAttemptTable} Comparison between the Corning SMF-28e+ fiber specification \cite{CorningSMF28} and BIFROST simulation of the same fiber. $S_0$ and $\lambda_0$ are the fitted values of the zero-dispersion slope and wavelength respectively.}
\renewcommand{\arraystretch}{1.3}
\begin{ruledtabular}
\begin{tabular}{lcc}
    \textbf{Parameter} & \textbf{Corning} & \textbf{Simulation} \\
    \hline
    $D_{\text{CD}}$, 1550 nm (ps/nm$\cdot$km) & $\leq 18$ & 12.31 \\
    $D_{\text{CD}}$, 1625 nm (ps/nm$\cdot$km) & $\leq 22$ & 16.15 \\
    $\lambda_0$ (nm) & 1304-1324 & 1350 \\
    $S_0$ (ps/(nm${}^2 \cdot$km)) & $\leq 0.092$ & 0.0749 \\
    $n_\text{eff}$, 1310 nm & 1.4674 & 1.4676 \\
    $n_\text{eff}$, 1550 nm & 1.4679 & 1.4680 \\
    Cutoff wavelength (nm) & $\leq 1260$ & 1318
\end{tabular}
\end{ruledtabular}
\end{table}

First, we note based on the table that several simulation results are within the Corning SMF-28e+ specifications, including the $D_{\text{CD}}$ at two wavelengths and the zero-dispersion slope $S_0$. The effective group refractive indices are also very close to specified values (within $0.02\%$). We take away an additional sign of agreement from Fig.~\ref{fig:CorningAttemptDCD}: the calculated $D_{\text{CD}} (\lambda)$ is fitted to the heuristic expression Eqn.~\ref{eqn:Heuristic_D_CD} and found to fit well, indicating that the simulated fiber has reasonable qualitative behavior in this regard. However, there are points of quantitative disagreement as well: the zero-dispersion wavelength is higher than the specified range, as is the cutoff wavelength, and the differences are rather significant. 

These comparisons indicate that the fibers we simulate are realistic in their chromatic dispersion properties: the simulated effective group refractive indices are well-matched, and the group velocity dispersion follows the expected behavior, providing some validation of the waveguide model and material models we use. There are areas of quantitative disagreement with the particular fiber we attempted to model, which we believe can be attributed to our lack of exact knowledge about the composition of and manufacturing methods for Corning SMF-28e+ fiber. Different manufacturing methods can add small amounts of additional dopants, such as chlorine, that slightly change the refractive indices and dispersion properties of the fiber \cite{Manufacturing_RefractiveIndex}; Corning may also deliberately use additional dopants not included here. Without this knowledge, achieving quantitative agreement is generally infeasible.

For the rest of this work, we use fiber inputs similar to the ones here, thus modeling a ``generic'' fiber whose dispersion properties are realistic.

\begin{figure}
\centering
\includegraphics[width=\columnwidth]{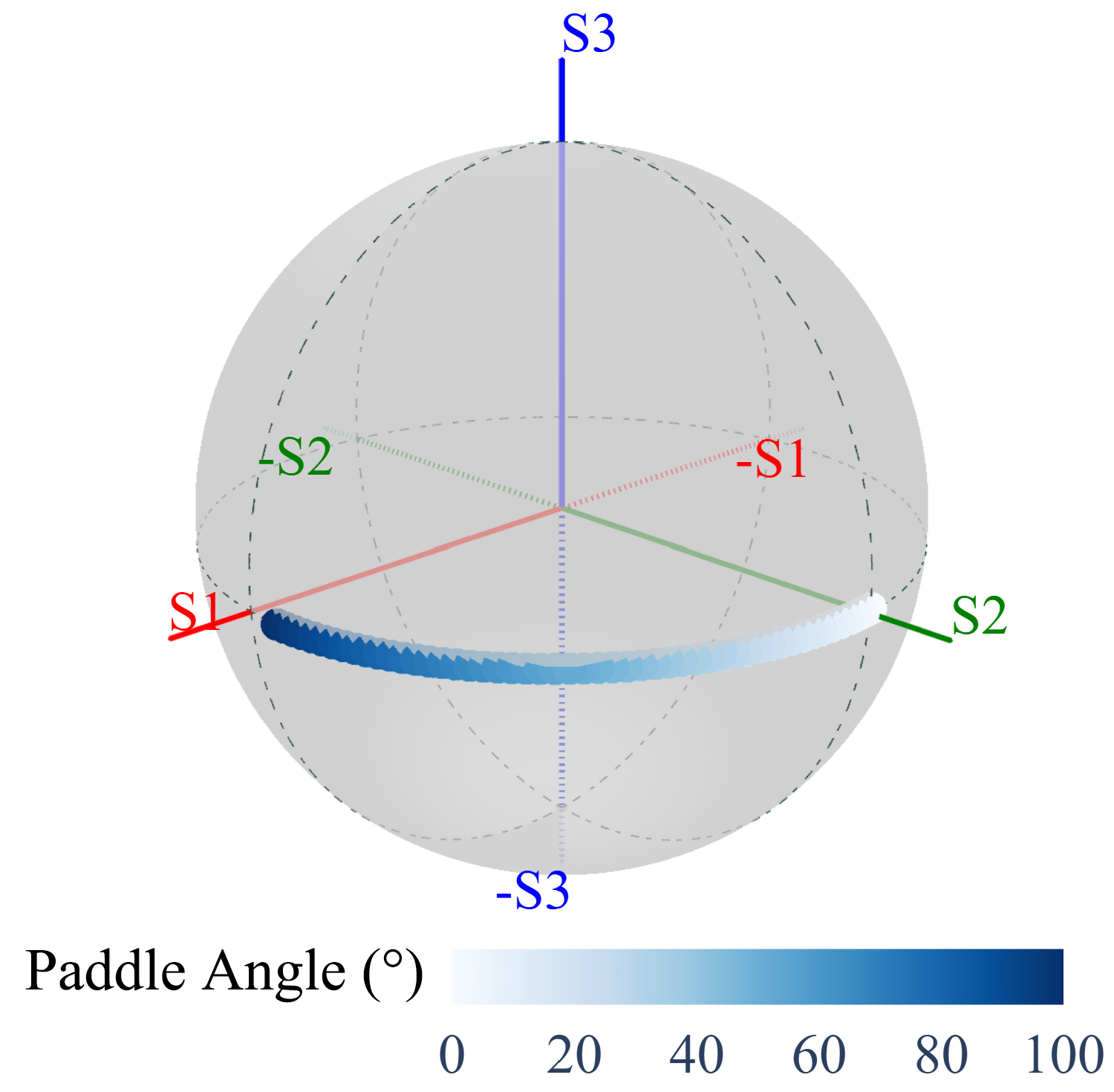}
\caption{\label{fig:PaddleValidation} Result of a simulation of a single fiber paddle with BIFROST. The input light at 1400~nm is $+45^\circ$ linearly polarized, and the fiber paddle is specified to have diameter 56~mm; six turns of fiber with cladding diameter $125~\mu$m is used to match the ThorLabs fiber paddle set manual \cite{ThorLabsPaddles}. The output polarization is shown here on the Poincar\'e sphere, with the color indicating the paddle angle. As the paddle is rotated, the polarization rotation is about the $S_3$ axis, corresponding to a rotation by a half-wave plate whose angle is being varied. This confirms that BIFROST reproduces the behavior specified in the ThorLabs fiber paddle manual. Figure partially made with PyPol \cite{PyPol}.}
\end{figure}

\paragraph{Validation of birefringence models and magnitudes.} In addition to verifying intrinsic fiber properties, it is important to verify the models of birefringence used in BIFROST. While all the equations we used are conventional models, each of them has limits; for instance, Eqn.~\ref{eqn:B_CNC} only works in the limit of small ellipticity, and Eqn.~\ref{eqn:B_BND} may fail when the bend radius is small (i.e. when the bend is sharp). Thus it would be useful to check that each model works within reasonable regimes and test the degree of their failure in extreme regimes. To test such models requires very simple fiber systems whose properties we know well. 

As an example of this type of validation, we consider fiber paddle systems such as ThorLabs FPC563. ThorLabs specifies \cite{ThorLabsPaddles} the total retardance of a single paddle given the paddle radius and the number of turns of fiber on the paddle: for instance, six turns of fiber with cladding diameter of $125~\mu$m wound onto a 56~mm diameter paddle should yield a retardance of half a wave at $\approx 1400$~nm. Fig.~\ref{fig:PaddleValidation} confirms that BIFROST reproduces this result. We simulated a single paddle with the above parameters, and we specified input light with $+45^\circ$ linear polarization at 1400~nm. The output polarization is shown on the Poincar\'e sphere as the paddle angle is varied, equivalent to turning a waveplate in free space \footnote{We also specify in the simulation that there is 2~cm of straight fiber length at the input of the paddle, which twists as the paddle is rotated; we ``measure'' the polarization right out of the loop. The twist rotates the birefringence axes relative to the input, which makes it physically similar to rotating a waveplate in free space.}. The resulting polarization rotation is almost exactly that expected of a rotating half-wave plate.

Such a test is a specific validation of the model we use for birefringence due to bending. Other systematic experimental observations are challenging to find but would be critically useful for verifying the different submodels implemented in BIFROST.

\paragraph{Validation of the top-level model.} In the past, the hinge model has successfully reproduced some statistical properties of the measured PMD of long installed fibers \cite{CzeglediOFC}. However, we adopt a different model of hinges, namely fiber paddles, which implies some restrictions on the kinds of rotations that can be caused by hinges in BIFROST. Additionally, we anticipate this model of alternating paddle hinges and long segments could model not only buried fiber (by keeping the long segments constant in time) but also aerial fiber, by allowing the long segments to vary with the environment -- this is work for the future.

One experimental technique that may be useful for this class of validation is polarization-sensitive optical time domain reflectometry (P-OTDR). Like regular OTDR, which indicates the locations in the fiber with high loss, P-OTDR is intended to measure the locations in the fiber with the highest PMD \cite{OG_POTDR, POTDR_1999, POTDR_2022}. This technique may produce results that help us validate the top-level model of BIFROST.

Of particular note in this category of validation, more clarity is needed regarding the limits of our model of spun fiber as discrete rotators. This model of spinning is able to reproduce the expected root-length scaling of PMD as well as the Maxwellian distribution of PMD statistics over random fiber ensembles (the latter will be shown in the next section); additionally, the spacing between the rotators is able to be chosen to empirically reproduce a PMD specification. However, the link between spin rate and effective rotator spacing is not clear, and the role of the amplitude of the sinusoidal spin rate function in this model is unclear. Moreover, the discretization means that the PMD of a given simulated fiber may depend strongly on the placement and rotations of the rotators; thus, when performing simulations, we should average over ensembles of fibers with different rotator realizations. This is an area of future work.

\paragraph{Comparisons to previous theoretical work.} As was discussed previously, there have been a number of previous attempts to model PMD in long fibers; some of these works (\cite[e.g.][]{OG_PMD_Analysis, HingeOG, Mabrouki_FiberModeling, Boudrioua_FiberModeling}) report on the specifics of their modeling choices and discuss specific birefringence mechanisms. In general, these works are not recent, and in the rare instances these works include comparison to experiment, the experiment is not well-documented. Nevertheless, comparisons between BIFROST and these previous results may show that BIFROST can reproduce previous computational results.

\begin{figure*}
\centering
    \includegraphics[width=0.45\textwidth]{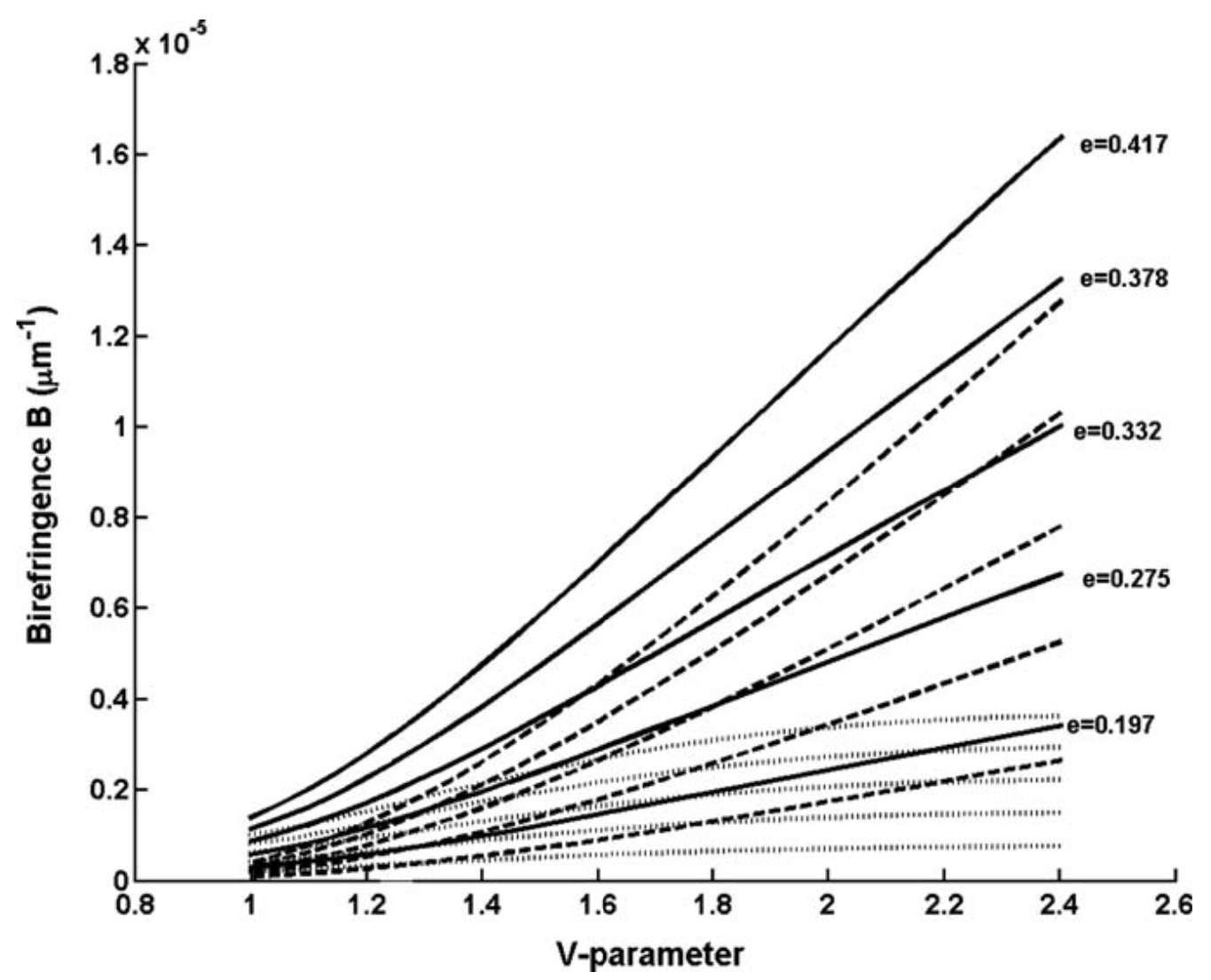}
    \includegraphics[width=0.45\textwidth]{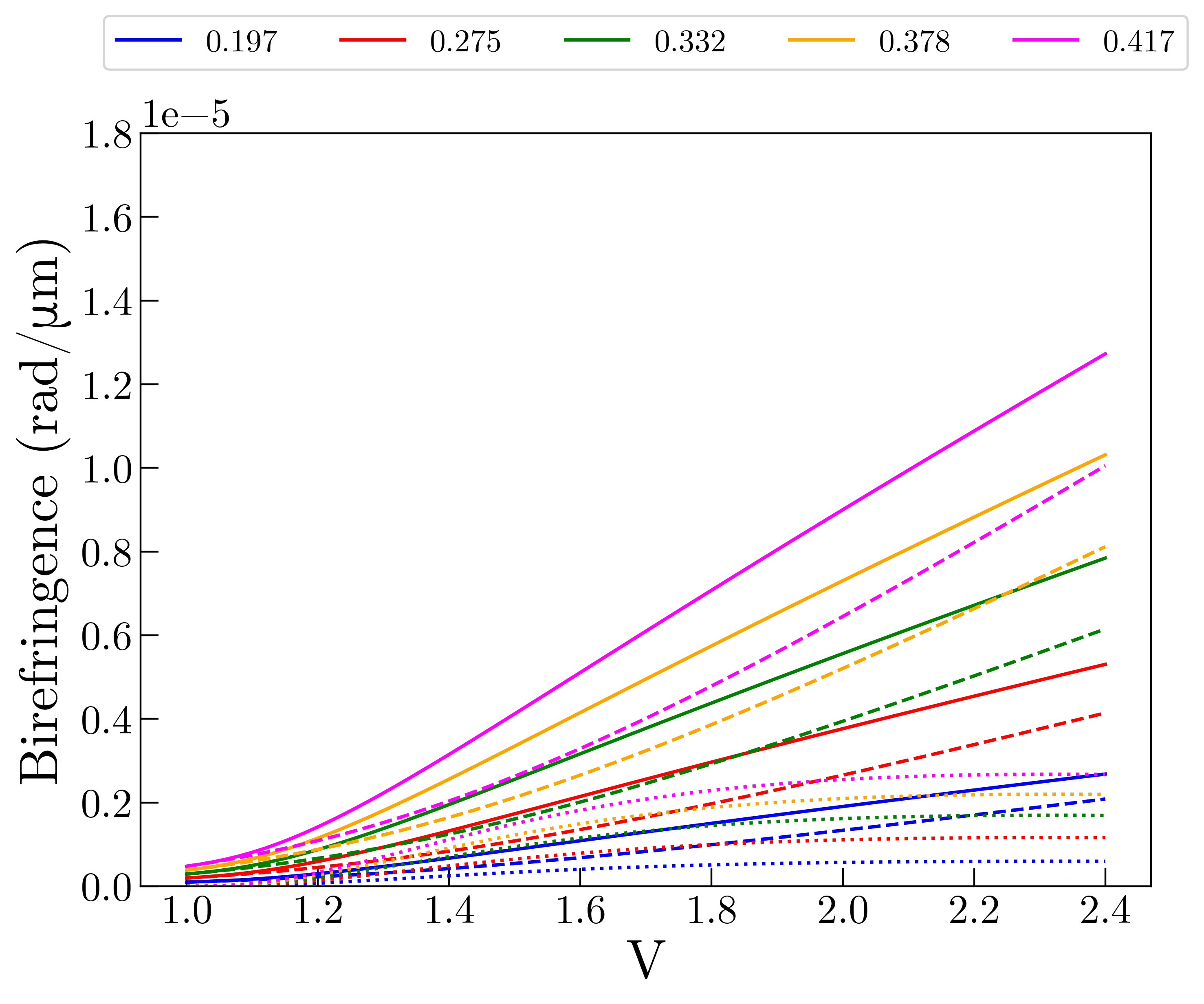}
    
    \includegraphics[width=0.45\textwidth]{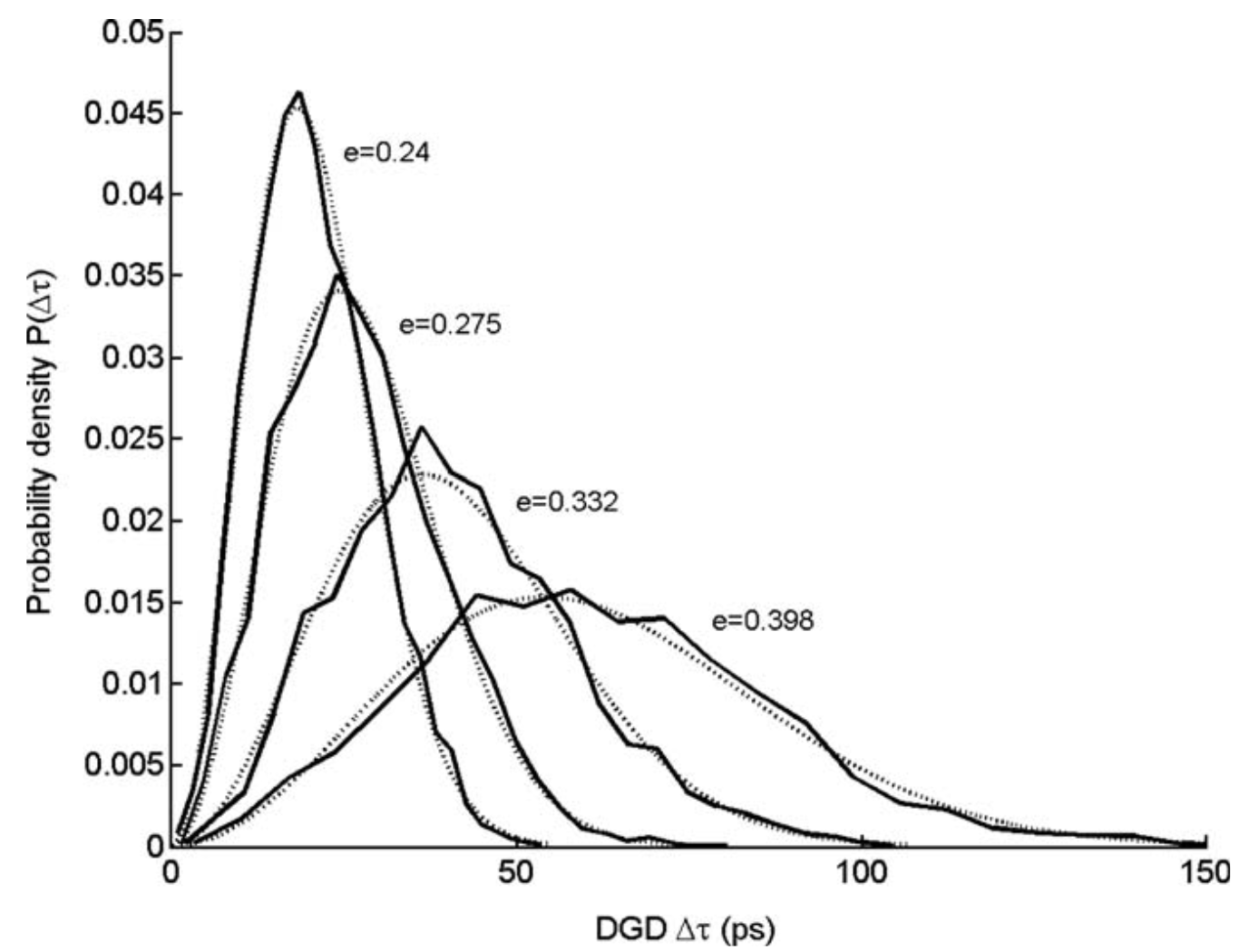}
    \includegraphics[width=0.45\textwidth]{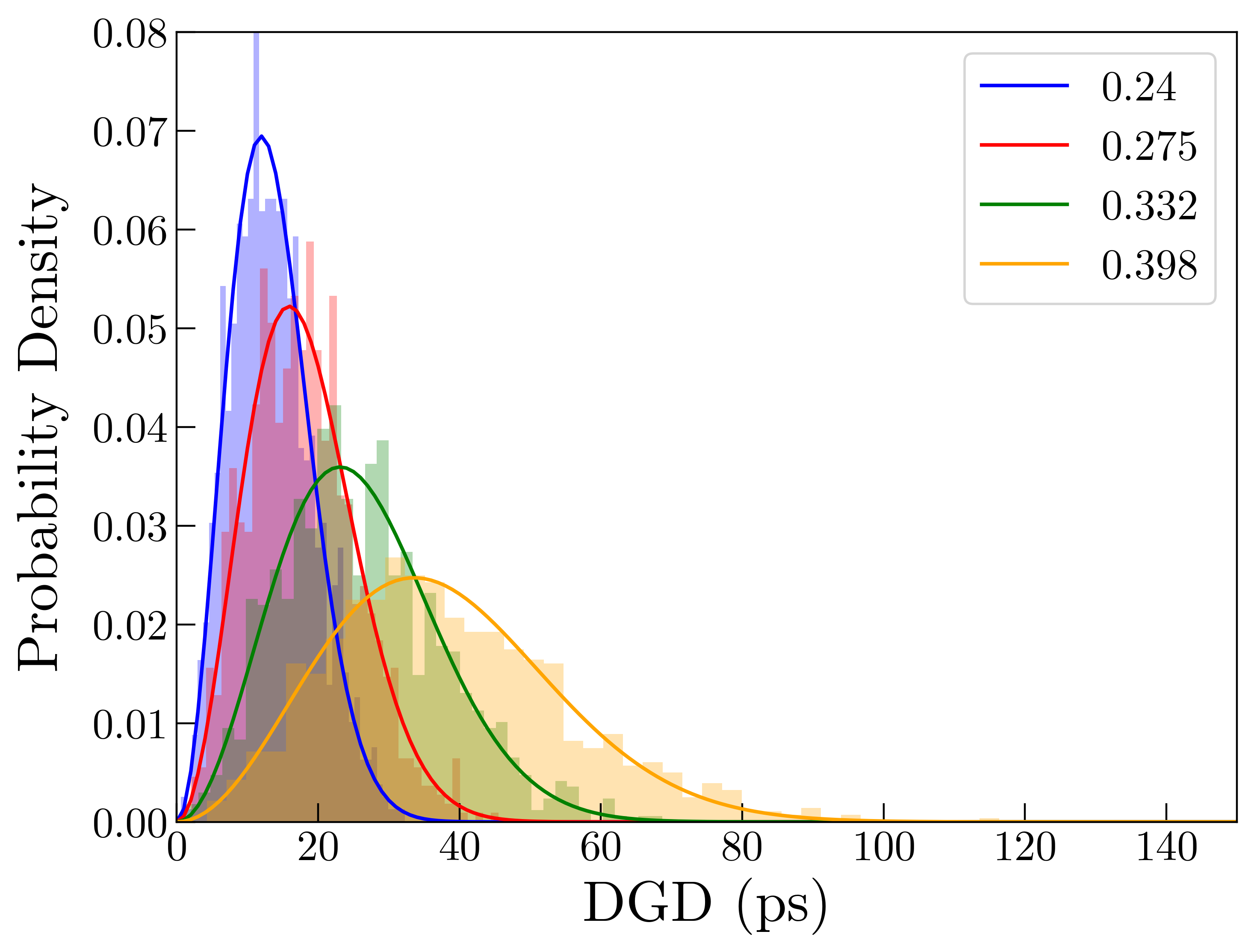}

\caption{\label{fig:GuptaComp} Comparison between the results of Ref.~\cite{OG_PMD_Analysis} (left panels) and calculations done by BIFROST (right panels). Top row: birefringence (in rad/$\mu$m) versus normalized frequency $V$ for several core eccentricities. The dotted curves are the birefringences due to core ellipticity alone, the dashed curves are birefringences due to asymmetric thermal stress alone, and the solid curves are the sum of the two for each eccentricity. Bottom row: Probability density of DGD as a function of core eccentricity, where the histograms are over random realizations of 800 arbitrary rotations spread across an 80~km fiber. The dotted curves on the left and solid curves on the right are fits to Maxwellian distributions. Qualitative agreement between Ref.~\cite{OG_PMD_Analysis} and BIFROST is clear; we believe that the visible quantitative discrepancy is due to internal inconsistency in Ref.~\cite{OG_PMD_Analysis}'s analysis, not to issues with BIFROST \cite{Note3}.}
\end{figure*}

Fig.~\ref{fig:GuptaComp} is an example of this type of validation. Here we compare results from BIFROST calculations to the work of Ref.~\cite{OG_PMD_Analysis}, a computational work that investigates birefringence as a function of core ellipticity. To make the comparison, we specify in BIFROST that our hinges should be arbitrary rotations (rather than fiber paddles); Ref.~\cite{OG_PMD_Analysis}'s model of PMD in a long fiber involves separating the fiber into many segments and randomly rotating the birefringence axes of each segment, which should produce similar statistical results to arbitrary rotator hinges. The top row of the figure shows birefringences due to core ellipticity and asymmetric thermal stress for a single straight segment. The bottom row shows the statistics of DGD for a 80~km fiber split into 800 segments of mean length 100~m. We randomly sample over many realizations of segment lengths and arbitrary rotations (with all other properties constant over all realizations) and histogram the resulting DGDs.

In both cases, qualitative agreement is achieved: the birefringences due to core ellipticity and asymmetric thermal stress show similar shapes as a function of the normalized frequency $V$, and the histograms from BIFROST fit well to Maxwellian curves just like the ones from Ref.~\cite{OG_PMD_Analysis}. However, there is quantitative disagreement of $\approx 30\%$ evident in both comparisons. We believe this discrepancy is mostly due to inconsistencies in Ref.~\cite{OG_PMD_Analysis} itself, not in BIFROST \footnote{As a consistency check, we substituted Ref.~\cite{OG_PMD_Analysis}'s material values into Ref.~\cite{OG_PMD_Analysis}'s own expression for the asymmetric thermal stress (Eqn. 3), and the resulting birefringence values differ from their own Figure 2, which is reproduced in the upper left panel in our Fig.~\ref{fig:GuptaComp}. The numbers we get this way vary from BIFROST's numbers by under 10\%, and the discrepancy is fully explained by slightly different values of material parameters. Given the internal numerical inconsistency in Ref.~\cite{OG_PMD_Analysis}, we are inclined to support BIFROST's calculations.}. We conclude that the comparisons of Fig.~\ref{fig:GuptaComp} point to good agreement, adding to the validity of BIFROST.

\paragraph{Statistical comparisons to previous experimental work.} Lastly, we may try to use BIFROST to directly simulate fibers whose PMD values are measured in experiments. Candidates for this type of comparison have been identified \cite{Validation_Cameron, PMDMeas1, PMDMeas2, PMDMeas_Quantum1, PMDMeas2_Quantum}. In all of these cases, the fiber is not so well-reported that the simulation parameters to be used in BIFROST are apparent. However, BIFROST's ability to analyze statistical ensembles allows us to explore the available parameter space and understand the parameter regimes under which BIFROST achieves qualitative and quantitative agreement with observations.
 
\section{BIFROST Usage and Recommended Operating Limits} 

As the previous section indicates, BIFROST is able to work with a wide variety of inputs. To simulate a given fiber, one must specify: the operating wavelength, the total length of the fiber, and the reference temperature for the measurement of the length; intrinsic fiber properties, namely the radii and doping of the core and cladding, the ellipticity of the core, and the effective spacing of arbitrary rotators (to model spun fiber); segment properties, namely the temperature, radius of curvature, and axial tension of each segment; and hinge properties, namely the number of paddles making up the hinge, the number of turns of fiber in each paddle, the radius and angle of each paddle, and the temperature of each paddle.

For a real-world installed fiber, it will be rare to know all of these parameters for the entire length of the fiber. Instead, BIFROST can be used to simulate the possible \emph{range} of behavior by simulating statistical ensembles of fibers, where the ensemble is defined by randomly drawing unknown or poorly known fiber properties from statistical distributions. For example, the user of a mostly buried fiber may not have access to one of the data closets to assess its temperature stability or see the configuration of the fiber inside; such a user can still use BIFROST to explore their questions of interest by simulating fibers with a variety of possible hinge geometries and temperatures. Of course, the effect of different hinge geometries and temperatures may itself be the the topic of interest. 

Besides simulating a real fiber, BIFROST's many required inputs forces theoretical studies of fiber PMD to specify operating regimes for each of the parameters and to consider the operating regime over which the study's conclusions are valid. Such studies will also likely use simulation over statistical ensembles to explore their question of interest over different parameter regimes.

The validation work above suggests that BIFROST can reproduce qualitatively reasonable behavior of fiber properties over a large parameter regime. Based on the limits of the approximations made and the validity range of the data used in BIFROST, we believe the codebase correctly computes supported contributions to birefringence in the following regime. 
\begin{itemize}
\item Single-mode operation, $V<2.405$
\item The weakly guiding regime $n_{\text{co}}-n_{\text{cl}} \ll 1$ (which implicitly requires weak germanium doping) (required by Eqn.~\ref{eqn:BetaFiberFinal})
\item The nearly-circular-core regime, $e^2 \ll 1$ (required by Eqn.~\ref{eqn:B_CNC})
\item Bend radii must be much larger than the cladding radius, $R \gg r_{\text{cl}}$ (required by Eqn.~\ref{eqn:B_BND})
\item Temperatures 200~K $\lesssim T \lesssim$ 300~K, limited by our model for the thermo-optic coefficient $dn/dT$ of bulk germania glass \cite{TempDependenceSiGe}. Our knowledge of the Sellmeier coefficients for germania glass is only at 297~K, but in the weakly doped regime, the temperature dependence of these coefficients is dominated by that of fused silica (which we know well).
\item Telecom wavelengths 1~$\mu$m $\lesssim \lambda \lesssim$ 2~$\mu$m. Our expression for the thermo-optic coefficient of bulk germania glass is measured at 1550~nm, but in the weakly doped regime, the core's refractive index is dominated by that of fused silica, which we know well over a broad range of wavelengths.
\end{itemize}
We do not model the temperature dependence of the coefficients of thermal expansion or the photoelastic constants $p_{11}$ and $p_{12}$ in fused silica and germania, as the variation is small within the above parameter regime. Other sources of birefringence omitted from BIFROST that could be consequential in some circumstances include cladding ellipticity, non-concentricity of core and cladding, external asymmetric (squeezing) stress, external transverse electric fields, and external longitudinal magnetic fields. The effects of polarization-dependent loss are also omitted in the present work.

In addition, we note that the model of spun fiber as discretized random rotations imposes a limit on the length scales of the behaviors that can be successfully modeled with BIFROST. Specifically, if the spacing between rotators is comparable to a length scale of interest (e.g. a disturbance of the fiber), such that the affected fiber length only contains a small number of rotators, then the resulting simulations are not likely to be accurate, and will instead strongly depend on the exact rotations and spacings. Analyses using BIFROST should be done in a regime where the length scales of interest are much larger than the rotator spacings, such that many rotations are included.

As a final usage note for BIFROST, modeling fibers so carefully makes BIFROST reminiscent of digital twin models \cite{DTReview1, DTReview2} for optical fiber networks. While digital twins are well-established for wireless communication \cite[e.g.][]{DTWireless}, it remains challenging to develop practical digital twins for fiber-based networks, and research in this area is ongoing \cite[e.g.][]{DT1, DT2, DT3}. For tracking state of polarization specifically, a digital twin is difficult to implement because the SOP at the end of the fiber is the accumulation of time-varying rotations along a fiber that result from several birefringence mechanisms. Parameters that could be used to predict the SOP would be challenging to acquire in deployed fiber. Thus digital twins are generally impractical for our purposes. Rather, BIFROST provides useful physics-based information about the \emph{statistical} performance of a fiber.

\section{Example Simulation: Wavelength-Division Multiplexing for PMD Compensation in Quantum Networks}

Within the operating regime described above, our validation work indicates that BIFROST produces realistic behaviors for simulated optical fibers, which makes BIFROST a powerful predictive tool. We demonstrate this for one specific research area of interest: PMD compensation schemes in fiber-based polarization-encoded quantum networks.

PMD represents a challenge for quantum networks not only because it changes the qubit states but also because the induced polarization rotation varies in time in ways that are challenging to predict, requiring an active stabilization system. PMD has been studied in a number of quantum network testbeds. In some of these testbeds, which we call ``quiet'' fiber links, PMD has contributed only a few percent entanglement-distribution fidelity error over timescales of several hours \cite{QunnectEntanglement, PMDMeas3_Quantum, QuietFiber3}. In other ``noisy'' testbeds, the fidelity reductions are more serious; for instance in Ref.~\cite{PMDMeas2_Quantum}, a 14~km fiber, buried except for $\approx 1$~km aerial, was analyzed every $\approx 15$~min and found to have drifted to 80\% process fidelity about 90\% of the time due to PMD drift. And in Ref.~\cite{PMDMeas_Quantum1}, which studied several fibers in the DC-QNET testbed, some fiber PMDs were found to contribute no more than a few percent infidelity over an hour, while other fibers were found to have process fidelities below 80\% after just one minute of PMD drift. Thus PMD is known to be a challenge for high-fidelity polarization-encoded quantum networks. (Besides the investigation of PMD compensation schemes we discuss below, BIFROST will also be useful in determining what factors contribute to these widely varying results, including whether the fiber is buried or aerial, fiber length, and whether the fiber is spun.) 

\begin{figure}
    \centering
    \includegraphics[width=0.7\columnwidth]{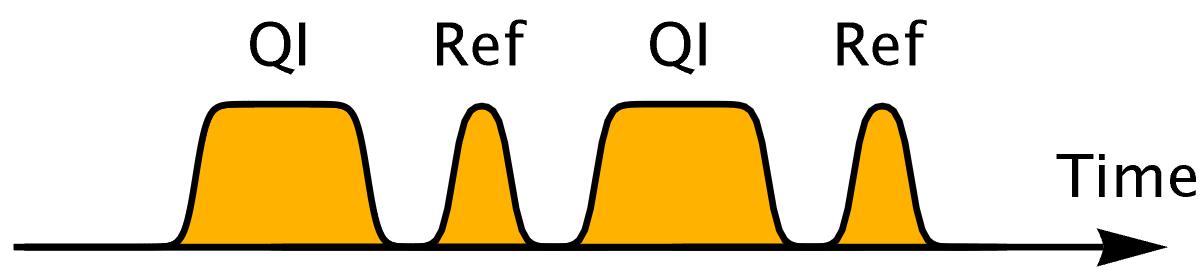} 

    \vspace{2.0em}
    \includegraphics[width=\columnwidth]{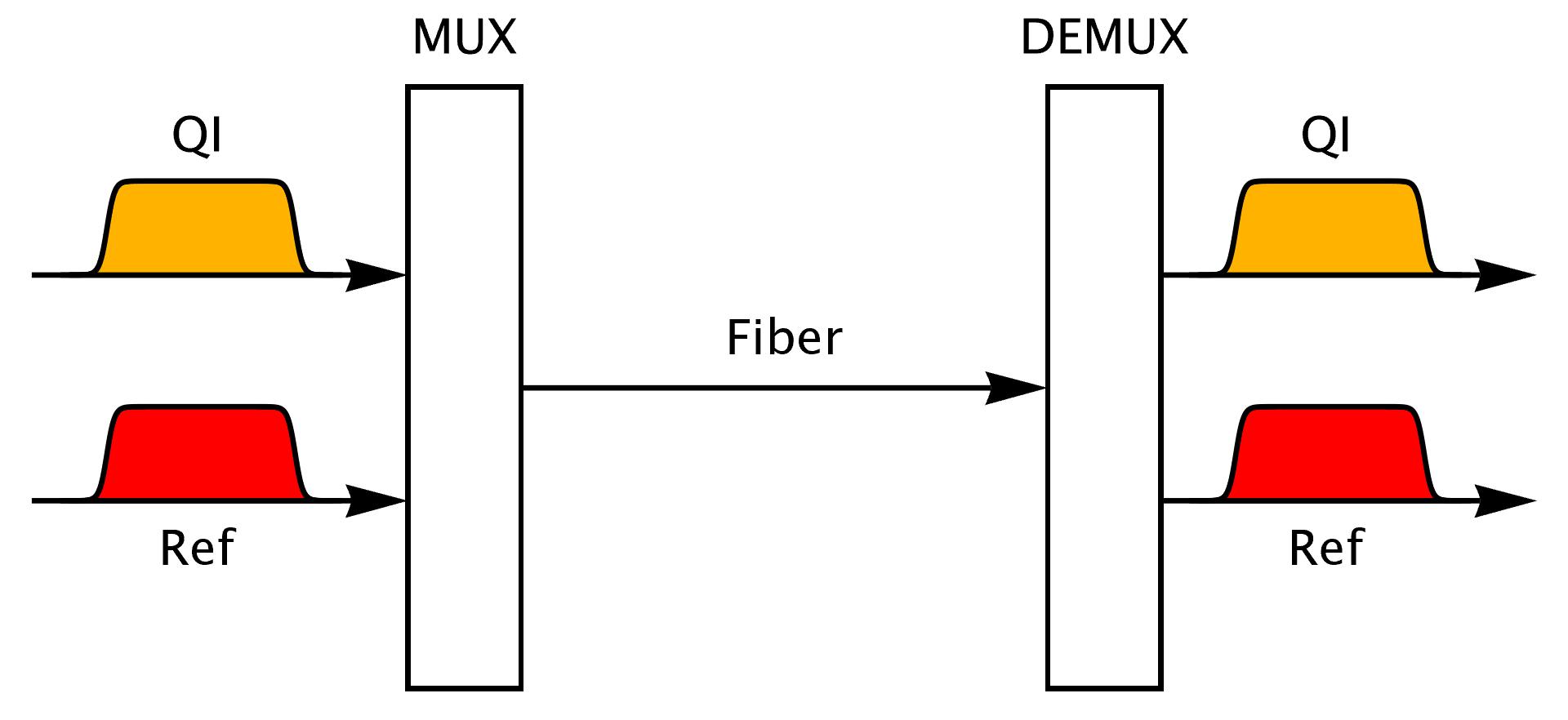} 
    
    \caption{\label{fig:CompensationCartoons} Common PMD compensation schemes. Top: time-division multiplexing (TDM), in which the quantum information (QI) is regularly paused long enough for reference light (Ref) at the same wavelength to measure the fiber PTF. Bottom: wavelength-division multiplexing (WDM), in which the quantum information and reference light (at a different wavelength) are multiplexed (MUX) into the same fiber and then demultiplexed (DEMUX) at the output.}
\end{figure}

Several PMD compensation schemes have been investigated and demonstrated. One class of compensation schemes, which has its roots in decades of work on networks dedicated to quantum key distribution (QKD), uses resources available in QKD (such as measuring the qubit error rate or using the discarded qubits) to perform compensation \cite{PMDComp_QKD1, PMDComp_QKD2, PMDComp_QKD3}. In more flexible quantum networks, which do not make any assumption about the kinds of tasks performed by the network \cite{QNReview_RMP_Azuma, QNReview_Science_RoadAhead}, compensation schemes  generally fall into two categories, as shown in Fig.~\ref{fig:CompensationCartoons}. One type is time-division multiplexing (TDM). In TDM, the quantum communication channel is paused long enough for classical reference pulses with known polarizations to be sent through the fiber. The polarization of these pulses is measured at the fiber output, and that information is used to correct the polarization rotations experienced by the quantum information. In between reference measurements, the network has no knowledge of the PMD drift, and the reference measurements and drift compensation must be updated regularly. TDM is  easy to implement , and  has been used for decades \cite[e.g.][]{PMD_TDM1, PMD_TDM2, PMD_TDM3, PMD_TDM4, PMD_TDM5, PMD_TDM6, PMD_TDM7, PMD_TDM8, PMD_TDM9}. However, it comes with the costs of network downtime and degraded fidelity in between measurements. (Ref.~\cite{QunnectEntanglement}, mentioned above, used TDM and demonstrated a network uptime of 99.83\% with average entanglement distribution fidelity of $\approx 95\%$, indicating that, for ``quiet'' fibers, TDM may produce good performance.)

An alternative to TDM is wavelength-division multiplexing (WDM). In WDM schemes, classical reference light at a wavelength different from the quantum channel is multiplexed into the same fiber as the quantum channel, and then demultiplexed at the output of the fiber and used to correct the quantum channel. Because of continuous copropagation, there is no network downtime or loss of fidelity from lack of measurement. But WDM faces two problems. First, light in the classical reference channel may scatter into other wavelengths, including the quantum channel, via nonlinear processes \cite{Agrawal, RecentProgress}. Second, as we have seen, fiber birefringence varies with wavelength, so the polarization rotation of each hinge and segment will vary, which gives rise to wavelength-dependent PMD. Uncontrolled time-variation of environmental parameters (e.g.  temperature and wind) further complicates matters so that PMD becomes increasingly uncorrelated between channels as the wavelength difference grows. Possibly as a result of these complications, few experiments have successfully demonstrated WDM-based PMD compensation. The few successes were largely under lab conditions (rather than using field-deployed fibers) and used small wavelength separations between the classical and quantum channels \cite{PMD_WDM1, PMD_WDM2, PMD_WDM3, PMD_WDM4}. In general, the practical utility of WDM for PMD compensation is an open question.

\begin{figure*}
\centering
    \includegraphics[width=0.9\textwidth]{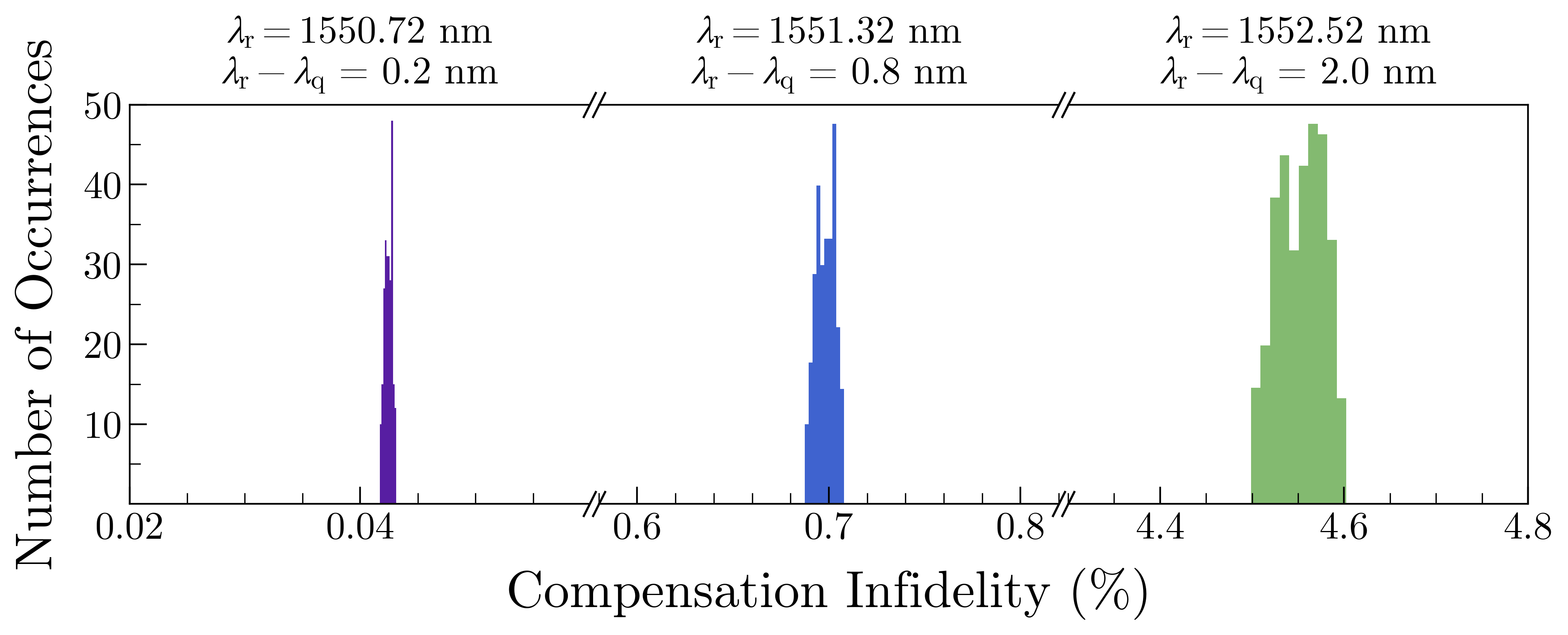}
\caption{\label{fig:WDMHist} Simulation of WDM compensation for three different reference wavelengths. The horizontal axis, which is broken in two places, shows the compensation infidelity $1-\mathcal{F}$. The histograms are the result of 250 random samples of hinge temperatures.}
\end{figure*}

As an example application of BIFROST, we simulate the effect of hinge temperature variation on WDM compensation for a mostly buried fiber link. Our simulated fiber is 26~km in length with five paddle hinges. Close to standard Corning SMF-28 fiber, we chose the core and cladding diameters as $8.2~\mu$m and $125~\mu$m, and the core doping as 3.6\% germania. The core ellipticity is chosen to be 0.5\%. The fiber we simulate is spun, with rotations inserted every 5~m. This results in fiber with average DGD of 0.204~ps, or 0.040~ps/km${}^{1/2}$, in line with the Corning spec of $<0.06$~ps/km${}^{1/2}$ for fiber PMD. Each hinge has three paddles with two, four, and two turns of fiber (chosen arbitrarily); the paddle radii are chosen from a normal distribution with mean and standard deviation 1~m and 0.1~m. This results in hinges with lengths of tens of meters. The paddle angles are allowed to take any value uniformly between $0^\circ$ and $360^\circ$. The paddle radius, number of turns and angle are selected at random for each paddle and held constant for the remainder of this simulation. The results presented below use a single member of this statistical ensemble. (Note: As discussed above, some statistics of the simulated fiber may depend strongly on the particular realization of rotators. We checked that the particular realization here has a total PMD near the ensemble average, a simple check that our sample fiber has no edge-case rotator choices.)

We imagine the five hinges to be in data closets or other spaces where the temperature may vary up to $2^\circ$C around a center set point of $20^\circ$C. Because the hinges are several kilometers apart, we can assume temperature changes at one hinge will be uncorrelated with those at the other hinges. In this simulation, we sample 250 sets of hinge temperatures from a uniform distribution and examine the PMD properties for each set of temperatures. BIFROST enables such explorations by being grounded in physical effects like temperature changes. By randomly sampling hinge temperatures, we can explore the impact of drifting hinge temperature on compensation fidelity.

To simulate WDM compensation, we model $\lambda_\text{Q} = 1550.52$~nm for the quantum channel (chosen for compliance with ITU standards) and the reference channel wavelength  $\lambda_\text{R}$. Denoting $J(\lambda)$ as the fiber Jones matrix PTF at wavelength $\lambda$, we compute the compensation fidelity as
\begin{equation}
    \mathcal{F}(\vec{V}) = \left| \vec{V} \cdot \left[ J^{-1}(\lambda_\text{R})J(\lambda_\text{Q}) \vec{V} \right] \right|^2.
\end{equation}
(Note that this is the fidelity for a single input state $\vec{V}$; it would perhaps be more useful to talk about an average fidelity over the space of possible states, but for simplicity we do not compute that here.)

The resulting compensation fidelity for horizontal input light is shown in Fig.~\ref{fig:WDMHist} for three values of $\lambda_\text{R}$. The horizontal axis plots compensation infidelity $1-\mathcal{F}$. The widths of the histograms indicate the range of compensation infidelities explored by the system over time (i.e. as the temperature varies). In this toy model, the histogram means are stationary and arise due to chromatic dispersion. These means could be trivially compensated in the case that they are stationary; however, real-world fiber exhibits more complex noise processes that give rise to drift over long timescales. (We anticipate that these processes can also be productively modeled with BIFROST, a possible direction of future work.)

The present WDM example indicates that, for a relatively simple fiber link and $\lambda_\text{R}-\lambda_\text{Q}=2$~nm, the compensation infidelity ranges between $\approx4.5\%$ and $\approx 4.6\%$. These results give an idea of how the compensation fidelity drops as $|\lambda_\text{R}-\lambda_\text{Q}|$ is increased. The simulations here also suggest that the WDM scheme can work for small enough $|\Delta \lambda|$, as also indicated by some experiments. For example, Ref.~\cite{PMD_WDM2} found that it was possible to compensate to better than 99\% fidelity using $\Delta\lambda = 0.8$~nm in a 8.5~km fiber with total DGD of 0.54~ps (this demonstration appears to have been done in a lab with a spooled fiber). 

This example simulation only scratches the surface of possible WDM compensation investigation, but it suggests that there are regimes in which WDM schemes may work well. It will be important to formulate more comprehensive metrics for compensation fidelity, investigate other sources of environmental variation, explore how the results change with the total PMD of the fiber, and compare WDM to TDM more directly to determine whether WDM is advantageous. Additionally, BIFROST may be useful in simulating more creative variations on WDM. For instance, Refs.~\cite{PMD_WDM3, PMD_WDM4} proposed using two reference wavelengths on opposite sides of the signal wavelength, and Ref.~\cite{Chapman} used heterodyne detection to relax power requirements on the WDM reference, resulting in negligible added noise from nonlinear scattering.

\section{Future Directions}

PMD compensation is one topic that can be studied with BIFROST. We envision other research directions and several extensions to BIFROST, detailed below.

\paragraph{New opportunities for validation.} As discussed, there are few published experimental works whose setups are documented in sufficient detail to permit a direct comparison to a BIFROST simulation, and there are few to no published works that experimentally test the individual contributions to PMD included in BIFROST, such as asymmetric thermal stress and torsional twisting. We look forward to greater opportunities for such comparison. 

In particular, we know of no tight link between experimental measurement of spinning-induced PMD and minimal, easy-to-compute numerical models for spinning. We adopted a model that uses discretized arbitrary rotations; this model reproduces some important observed properties of spun fiber PMD, but it also has some caveats to be investigated, and improvements to the model are the subject of future work. We note that the lack of a good model of spun fiber PMD is made more visible through the work of developing a unified model of fiber polarization mode dispersion, as we have done here.

We also find few works attempting to reverse engineer the composition, geometry, and manufacturing variation of proprietary commercial fiber. The modeling of dispersion-compensated fiber would be an especially useful addition to the toolkit. 

The historical lack of a systematic structure for unifying the physical, optical, and material properties of fiber made it hard to see gaps in the literature. Some are now evident to us. We hope others are motivated to fill some of these gaps and that BIFROST can serve as an open-source scaffold for collecting and combining such models. 

But we also note that some of the parameters needed for a BIFROST simulation may not be able to be determined, such as the compositions of commercial fibers, which may be trade secrets. Thus, exploration is needed even to develop testable hypotheses that could be experimentally checked with installed fibers. This will require broad exploration of the behavior of PMD or output SOP with various fiber and environmental characteristics.

\paragraph{Developing intuition by simulating simple systems.} The simulations of BIFROST are challenging to understand intuitively. A research direction that might be fruitful is the simulation of simple fiber systems, e.g. a single fiber length with a single hinge on the end, or a single hinge in between a pair of fiber segments. Investigating these systems as functions of environmental variables like temperature might help us develop intuition about installed fibers, the hierarchy of birefringences typical in these fibers, and the kinds of fiber configurations that will be better or worse for polarization-sensitive applications. For instance, we might ask questions about how well hinges need to be controlled to make a fiber quiet enough for TDM schemes with acceptable downtime.

\paragraph{Optimizing TDM with a second reference wavelength.} Data taken in an installed fiber between the University of Maryland and Army Research Labs campuses suggests that, while the output state of polarization at two different wavelengths is very uncorrelated when the wavelengths are widely separated (more than a few nm), the speed of the drift in output SOP may be correlated. This raises the question of whether a reference wavelength distinct from the quantum channel could be used to optimize a TDM scheme, the idea being to monitor the polarization state of the reference wavelength and, when its SOP drifts too far as defined by a user, pause the quantum channel and perform TDM compensation. This scheme might allow a network to stop quantum communication less often to do TDM if there are times when the fiber is relatively quiet, and do it more often if the fiber is relatively noisy. (For instance, we might perform compensation less often at night, when PMD has often been observed to drift more slowly \cite{PMDMeas_Quantum1, PMDMeas2_Quantum}.) The possible gain in uptime from such a scheme depends strongly on the speed of PMD drifts in the fiber and the threshold for stopping to do TDM, which relates to the desired fidelity in the system. Such schemes could be simulated with BIFROST.

\paragraph{Including nonlinear effects and polarization-dependent loss.} There are a number of phenomena in optical fibers that are not included in BIFROST but that may be useful for some applications. One is nonlinear scattering \cite{Agrawal, RecentProgress}; such scattering is of interest for instance in WDM compensation schemes, where the bright reference channel may scatter photons into the quantum channel, thereby introducing noise in this channel \cite{Coex1, Coex2}. Another phenomenon is polarization-dependent loss, which has already been investigated in the context of entanglement distribution in quantum networks \cite{ESD_PDL}.

\paragraph{Simulating correlations between members of a Monte Carlo ensemble.} It is often the case that many optical fibers are installed together in a cable. Members of such a set experience common environmental fluctuations, with the result that PMD fluctuations between fiber strands in a bundle may be well-correlated \cite{PMDMeas1}. This is a natural area of exploration for BIFROST, which could compare members of a statistical ensemble of fibers and explore correlations in PMD fluctuations across the set. The possibility of using one fiber for a reference in PMD compensation while using another for the quantum channel is particularly intriguing.

\paragraph{Simulating entanglement distribution fidelity.} Some quantum network tasks require that two entangled photons are sent to two different locations over two different fibers, or two different sources produce photons to be sent to a common source over two different fibers. The two fibers will each have their own environmental fluctuations and PMD drifts. How possible is it to preserve entanglement fidelity in this situation, and what kinds of compensation will be needed? While some previous work has investigated these questions \cite{ESD_PMD1, ESD_PMD2, ESD_PMD3, ESD_PDL}, these remain open questions that BIFROST could help researchers understand.

\paragraph{Extending beyond single-core single-mode silica fibers.} One possibility that quantum-network researchers may have to contend with is that, because of PMD in deployed fibers, quantum networks may never reach the performance required for some tasks. If that is the case, considerations outside of this paradigm are worthwhile. One direction is, of course, using different information encodings in the photons, such as time-bin or frequency-bin encoding. 

Another approach is to use different types of fiber, two of which we mention here. One type of fiber that may be promising is multi-core fiber \cite{Multicore1, Multicore2, Multicore3}, which may be advantageous because the cores will experience common environments and stresses, so their PMD drifts would be highly correlated. This would enable compensation schemes where one core is used for reference light and another is used for quantum communication. However, the multiple cores also have crosstalk. Another type of fiber that may be useful in quantum networking is hollow-core fibers, whose core is air \cite{Hollowcore1, Hollowcore2}. The waveguiding proceeds by effects other than total internal reflection, but because air has very little birefringence, these fibers have the potential to enable high-fidelity quantum networks.

Despite the possible advantages, the installation of new fiber for quantum applications is likely to be costly; extending BIFROST to include these kinds of fibers could help industries and governments determine if installing new fiber is worth the cost.

\section{Conclusion}

We have developed a first-principles model of polarization mode dispersion in long optical fibers. The model uses analytical results for step-index cylindrical waveguides, data from measurements of silica-germania binary glasses, and models of several physical birefringence mechanisms to calculate total Jones matrices for long fibers. Spun fibers are modeled via arbitrary rotations along birefringent fiber lengths, while hinges modeled as sets of fiber paddles allow for the simulation of buried fibers with short lengths above ground. We discussed and began to perform validation of the model and then demonstrated the model's utility by exploring an open problem, the feasibility of wavelength-division multiplexing PMD compensation schemes in quantum networks. 

Compared to previous models, BIFROST focuses on physical principles, allowing researchers to gain understanding of the primary PMD mechanisms in their fiber links. In contexts where PMD is useful, such as in some fiber sensing applications, BIFROST may be able to help researchers find new use cases or investigate sensitivity. In contexts where PMD should be mitigated, such as quantum networking, BIFROST may help researchers determine the sources of PMD in their networks, investigate best practices for mitigation and compensation, and make decisions on investments into fiber stability. BIFROST bridges the gap between prior work, in which telecommunications needs motivated the discovery of many insights about PMD in long fibers, and current research domains, such as fiber sensors and quantum networks. By providing a first-principles model of PMD and implementing it in an open-source Python library \cite{Git}, we expect this model to be a useful framework for researchers studying a wide array of fiber-based technologies.

\vspace{1em}
\noindent {\it Acknowledgments.---} We are grateful to Deven Bowman and Evan McClintock for generating useful insights leading to this project. We also thank Deniz Kurdak, Yaxin Li, and Trey Porto for additional discussions and activity leading up to this work. This work was supported by MAQP in partnership with the CCDC Army Research Laboratory (ARL) W911NF2420107. P.R.B. also acknowledges support from the ARCS Foundation Metro-Washington Chapter. The mention of specific companies or products does not imply an endorsement of these products by the University of Maryland, ARL, or the ARCS Foundation.

\bibliography{main.bib}

@PREAMBLE{
 "\providecommand{\noopsort}[1]{}" 
 # "\providecommand{\singleletter}[1]{#1}%" 
}

@misc{Git,
  author = {Patrick Banner},
  title = {{BIFROST}},
  howpublished = {GitHub},
  year = {2025},
  note = "\url{https://https://github.com/JQIamo/BIFROST/}"
}

@manual{ThorLabsPaddles,
  title = {{Manual Fiber Polarization Controllers: User Guide}},
  address = {Newton, NJ},
  note = {Rev. M, available at \url{https://www.thorlabs.com/drawings/db208e0c180e1283-5E0F377D-C780-4DD8-1470FB2A1D365C40/FPC563-Manual.pdf}},
  organization = {ThorLabs},
  year = {2024}
}

@manual{CorningSMF28,
  title = {{Corning SMF-28e+ Optical Fiber Product Information}},
  address = {Corning, NY},
  note = {Rev. PI1463, available at \url{https://www.corning.com/media/worldwide/coc/documents/Fiber/product-information-sheets/PI-1463-AEN.pdf}},
  organization = {Corning},
  year = {2024}
}

@article{Chraplyvy.Review, 
    year = {2018}, 
    title = {{Fiber-optic transmission and networking: the previous 20 and the next 20 years [Invited]}}, 
    author = {Winzer, Peter J and Neilson, David T and Chraplyvy, Andrew R}, 
    journal = {Optics Express}, 
    doi = {10.1364/oe.26.024190}, 
    pages = {24190}, 
    number = {18}, 
    volume = {26}
}

@article{Savory_Review, 
    year = {2008}, 
    title = {{Digital filters for coherent optical receivers}}, 
    author = {Savory, Seb J}, 
    journal = {Optics Express}, 
    doi = {10.1364/oe.16.000804},
    pages = {804}, 
    number = {2}, 
    volume = {16}
}

@article{QNReview_RMP_Azuma, 
    year = {2023}, 
    title = {{Quantum repeaters: From quantum networks to the quantum internet}}, 
    author = {Azuma, Koji and Economou, Sophia E. and Elkouss, David and Hilaire, Paul and Jiang, Liang and Lo, Hoi-Kwong and Tzitrin, Ilan}, 
    journal = {Reviews of Modern Physics}, 
    issn = {0034-6861}, 
    doi = {10.1103/revmodphys.95.045006}, 
    pages = {045006}, 
    number = {4}, 
    volume = {95}
}

@article{QNReview_Science_RoadAhead, 
    year = {2018}, 
    title = {{Quantum internet: A vision for the road ahead}}, 
    author = {Wehner, Stephanie and Elkouss, David and Hanson, Ronald}, 
    journal = {Science}, 
    issn = {0036-8075}, 
    doi = {10.1126/science.aam9288}, 
    number = {6412}, 
    volume = {362}
}

@article{AtomicClocksReview,
  title = {Optical atomic clocks},
  author = {Ludlow, Andrew D. and Boyd, Martin M. and Ye, Jun and Peik, E. and Schmidt, P. O.},
  journal = {Reviews of Modern Physics},
  volume = {87},
  issue = {2},
  pages = {637--701},
  numpages = {65},
  year = {2015},
  doi = {10.1103/RevModPhys.87.637}
}

@article{EarlyPMD1, 
    year = {1978}, 
    title = {{Polarization mode dispersion in single-mode fibers}}, 
    author = {Rashleigh, S. C. and Ulrich, R.}, 
    journal = {Optics Letters}, 
    doi = {10.1364/ol.3.000060},  
    pages = {60}, 
    number = {2}, 
    volume = {3}
}

@article{EarlyPMD2, 
    year = {1986}, 
    title = {{Phenomenological approach to polarisation dispersion in long single-mode fibres}}, 
    author = {Poole, C D and Wagner, R E}, 
    journal = {Electronics Letters}, 
    doi = {10.1049/el:19860703}, 
    pages = {1029}, 
    number = {19}, 
    volume = {22}
}

@article{Mochizuki, 
    year = {1981}, 
    title = {{Polarisation mode dispersion measurements in long single mode fibres}}, 
    author = {Mochizuki, K and Namihira, Y and Wakabayashi, H}, 
    journal = {Electronics Letters}, 
    doi = {10.1049/el:19810108}, 
    pages = {153}, 
    number = {4}, 
    volume = {17}
}

@article{EarlyPMDTheory1, 
    year = {1991}, 
    title = {{Statistical theory of polarization dispersion in single mode fibers}}, 
    author = {Foschini, G.J. and Poole, C.D.}, 
    journal = {Journal of Lightwave Technology}, 
    doi = {10.1109/50.97630}, 
    pages = {1439--1456}, 
    number = {11}, 
    volume = {9}
}

@article{EarlyPMDTheory2, 
    year = {1991}, 
    title = {{Polarization mode dispersion of short and long single-mode fibers}}, 
    author = {Gisin, N. and Weid, J.-P. Von der and Pellaux, J.-P.}, 
    journal = {Journal of Lightwave Technology}, 
    doi = {10.1109/50.85780}, 
    pages = {821--827}, 
    number = {7}, 
    volume = {9}
}

@article{SpinningOG,
    author = {A. J. Barlow and J. J. Ramskov-Hansen and D. N. Payne},
    journal = {Applied Optics},
    number = {17},
    pages = {2962--2968},
    title = {Birefringence and polarization mode-dispersion in spun single-mode fibers},
    volume = {20},
    year = {1981},
    doi = {10.1364/AO.20.002962}
}

@inproceedings{SpinningLongFiber,
    author = {D. Sarchi and G. Roba},
    booktitle = {Optical Fiber Communication Conference},
    journal = {Optical Fiber Communication Conference},
    pages = {WJ2},
    publisher = {Optica Publishing Group},
    title = {PMD Mitigation Through Constant Spinning and Twist Control: Experimental Results},
    year = {2003},
    city = {Atlanta, GA, USA}
}

@inproceedings{SpinningHistoryAndTheory,
    author = {Ming-Jun Li and Xin Chen and Daniel A. Nolan},
    title = {{Fiber spinning for reducing polarization mode dispersion in single-mode fibers: theory and applications}},
    volume = {5247},
    booktitle = {Optical Transmission Systems and Equipment for WDM Networking II},
    editor = {Benjamin B. Dingel and Werner Weiershausen and Achyut K. Dutta and Ken-Ichi Sato},
    organization = {International Society for Optics and Photonics},
    publisher = {SPIE},
    pages = {97 -- 110},
    year = {2003},
    doi = {10.1117/12.512063},
    city = {Orlando, FL, USA}
}

@article{SpinningTheory2,
    author = {A. Galtarossa and M. Guglielmucci and L. Palmieri and L. Schenato and Carlo G. Someda and},
    title = {Modeling and Design of Low-PMD Spun Fibers},
    journal = {Fiber and Integrated Optics},
    volume = {27},
    number = {4},
    pages = {216--222},
    year = {2008},
    doi = {10.1080/01468030802191841}
}

@article{SpinningNonlocal,
    title = {{Anisotropy in spun single-mode fibres}},
    author = {A. J. Barlow and J. J. Ramskov-Hansen and D. N. Payne},
    journal = {Electronics Letters},
    volume = {18},
    number = {5},
    doi = {10.1049/el:19820138},
    year = {1982}
}

@article{SpinningTheoryZeros2019,
    author = {Sad{\i}k, \c{S}. A. and Karl{\i}k, S. E. and Temurta\c{s}, H. and Altuncu, A.},
    year = {2019},
    title = {{Optimization of spin parameters for spun fibers having low polarization mode dispersion}},
    journal = {Optical Fiber Technology},
    volume = {51},
    pages = {31-–35},
    doi = {10.1016/j.yofte.2019.04.011}
}

@article{Buildings1,
    title = {Application of optical fiber distributed sensing to health monitoring of concrete structures},
    journal = {Mechanical Systems and Signal Processing},
    volume = {39},
    number = {1},
    pages = {441-451},
    year = {2013},
    doi = {10.1016/j.ymssp.2012.01.027},
    author = {Sergi Villalba and Joan R. Casas}
}

@Article{Buildings2,
    AUTHOR = {Bado, Mattia Francesco and Casas, Joan R.},
    TITLE = {A Review of Recent Distributed Optical Fiber Sensors Applications for Civil Engineering Structural Health Monitoring},
    JOURNAL = {Sensors},
    VOLUME = {21},
    YEAR = {2021},
    NUMBER = {5},
    ARTICLE-NUMBER = {1818},
    DOI = {10.3390/s21051818}
}

@article{LongBaseline1,
  title = {{Longer-Baseline Telescopes Using Quantum Repeaters}},
  author = {Gottesman, Daniel and Jennewein, Thomas and Croke, Sarah},
  journal = {Physical Review Letters},
  volume = {109},
  issue = {7},
  pages = {070503},
  year = {2012},
  doi = {10.1103/PhysRevLett.109.070503}
}

@article{InterferometryAstro,
author = {John D. Monnier},
year = {2003},
journal = {Reports on Progress in Physics},
volume = {66},
number = {5},
pages = {789},
doi = {10.1088/0034-4885/66/5/203}
}

@article{DCF1,
    title = {{Dispersion Compensating Fibers}},
    journal = {Optical Fiber Technology},
    volume = {6},
    number = {2},
    pages = {164--180},
    year = {2000},
    doi = {https://doi.org/10.1006/ofte.1999.0324},
    author = {Lars Gr\"uner-Nielsen and Stig Nissen Knudsen and Bent Edvold and Torben Veng and Dorte Magnussen and C. Christian Larsen and Hans Damsgaard}
}

@article{DCF2,
    author = {Lars Gr\"{u}ner-Nielsen and Marie Wandel and Poul Kristensen and Carsten Jorgensen and Lene Vilbrad Jorgensen and Bent Edvold and Bera P\'{a}lsd\'{o}ttir and Dan Jakobsen},
    journal = {Journal of Lightwave Technology},
    number = {11},
    pages = {3566},
    title = {{Dispersion-Compensating Fibers}},
    volume = {23},
    year = {2005}
}

@article{QuantumNetworkFibers_RMP,
  title = {Colloquium: Cavity-enhanced quantum network nodes},
  author = {Reiserer, Andreas},
  journal = {Reviews of Modern Physics},
  volume = {94},
  issue = {4},
  pages = {041003},
  numpages = {22},
  year = {2022},
  doi = {10.1103/RevModPhys.94.041003}
}

@article{PMDMeas_Quantum1,
    author = {McKenzie, Wayne and Richards, Anne Marie and Patel, Shirali and Gerrits, Thomas and Akin, T. G. and Peil, Steven and Black, Adam T. and Tulchinsky, David and Hastings, Alexander and Li-Baboud, Ya-Shian and Rahmouni, Anouar and Burenkov, Ivan A. and Mink, Alan and Diaz, Matthew and Lal, Nijil and Shi, Yicheng and Kuo, Paulina and Shrestha, Pranish and Merzouki, Mheni and Rodriguez Perez, Alejandro and Onuma, Eleanya and Jones, Daniel E. and Davis, Atiyya A. and Searles, Thomas A. and Whalen, J. D. and Quraishi, Qudsia Sara and Collins, Kate S. and Cooper, La Vida and Shaw, Harry and Crabill, Bruce and Slattery, Oliver and Battou, Abdella},
    title = {{Clock synchronization characterization of the Washington DC metropolitan quantum network (DC-QNet)}},
    journal = {Applied Physics Letters},
    volume = {125},
    number = {16},
    pages = {164004},
    year = {2024},
    doi = {10.1063/5.0225082}
}

@article{PMDMeas2_Quantum,
  title={{Demonstration of quantum network protocols over a 14-km urban fiber link}},
  author={Kucera, Stephan and Haen, Christian and Arensk{\"o}tter, Elena and Bauer, Tobias and Meiers, Jonas and Sch{\"a}fer, Marlon and Boland, Ross and Yahyapour, Milad and Lessing, Maurice and Holzwarth, Ronald and Becher, Christoph and Eschner, J\"urgen},
  journal={npj Quantum Information},
  volume={10},
  number={1},
  pages={88},
  year={2024}
}

@article{PMDMeas3_Quantum,
    author = {M. A. Zalewski and D. Wu and A. L. Ferrari and Y. Xie and N. M. Linke},
    title = {{Kilometer-Scale Ion-Photon Entanglement with a Metastable ${}^{88}$Sr$^{+}$ Qubit}},
    journal = {ArXiv},
    year = {2025},
    doi = {10.48550/arXiv.2506.11257}
  
}

@article{PMDComp_QKD1, 
    year = {2017}, 
    title = {{Polarization-basis tracking scheme for quantum key distribution using revealed sifted key bits}}, 
    author = {Ding, Yu-Yang and Chen, Wei and Chen, Hua and Wang, Chao and li, Ya-Ping and Wang, Shuang and Yin, Zhen-Qiang and Guo, Guang-Can and Han, Zheng-Fu}, 
    journal = {Optics Letters}, 
    doi = {10.1364/ol.42.001023}, 
    pages = {1023}, 
    number = {6}, 
    volume = {42}
}

@article{PMDComp_QKD2, 
    year = {2021}, 
    title = {{Fibre polarisation state compensation in entanglement-based quantum key distribution}}, 
    author = {Shi, Yicheng and Poh, Hou Shun and Ling, Alexander and Kurtsiefer, Christian}, 
    journal = {Optics Express}, 
    doi = {10.1364/oe.437896}, 
    pages = {37075}, 
    number = {23}, 
    volume = {29}
}

@article{PMDComp_QKD3, 
    year = {2018}, 
    title = {{Field implementation of long-distance quantum key distribution over aerial fiber with fast polarization feedback}}, 
    author = {Li, Dong-Dong and Gao, Song and Li, Guo-Chun and Xue, Lu and Wang, Li-Wei and Lu, Chang-Bin and Xiang, Yao and Zhao, Zi-Yan and Yan, Long-Chuan and Chen, Zhi-Yu and Yu, Gang and Liu, Jian-Hong}, 
    journal = {Optics Express}, 
    doi = {10.1364/oe.26.022793}, 
    pages = {22793}, 
    number = {18}, 
    volume = {26}
}

@article{QuietFiber3,
    title = {{Passively stable distribution of polarisation entanglement over 192 km of deployed optical fibre}},
    author = {S\"oren Wengerowsky and Siddarth Koduru Joshi and Fabian Steinlechner and Julien R. Zichi and Bo Liu and Thomas Scheidl and Sergiy M. Dobrovolskiy and René van der Molen and Johannes W. N. Los and Val Zwiller and Marijn A. M. Versteegh and Alberto Mura and Davide Calonico and Massimo Inguscio and Anton Zeilinger and Andr\'e Xuereb and Rupert Ursin},
    journal = {npj Quantum Information},
    volume = {6},
    number = {5},
    year = {2020},
    doi = {10.1038/s41534-019-0238-8}
}

@article{PMD_TDM1, 
year = {2006}, 
title = {{Polarization recovery and auto-compensation in quantum key distribution network}}, 
author = {Ma, Lijun and Xu, Hai and Tang, Xiao}, 
journal = {Quantum Communications and Quantum Imaging IV}, 
doi = {10.1117/12.679575}, 
pages = {630513-1--630513-6},
city = {San Diego, CA, USA},
publisher = {SPIE}
}

@article{PMD_TDM2, 
year = {2007}, 
title = {{Experimental demonstration of an active quantum key distribution network with over gbps clock synchronization}}, 
author = {Ma, Lijun and Mink, A. and Xu, Hai and Slattery, O. and Tang, Xiao}, 
journal = {IEEE Communications Letters}, 
doi = {10.1109/lcomm.2007.071477}, 
pages = {1019--1021}, 
number = {12}, 
volume = {11}
}

@article{PMD_TDM3, 
year = {2015}, 
title = {{Towards polarisation-encoded quantum key distribution in optical fibre networks}}, 
author = {Pillay, Sharmini and Mirza, Abdul and Petruccione, Francesco}, 
journal = {South African Journal of Science}, 
doi = {10.17159/sajs.2015/20130380}, 
number = {7/8}, 
volume = {111}
}

@article{PMD_TDM4, 
year = {2015}, 
title = {{Adaptive Polarization-State Monitoring and Stabilization Scheme for One-Way Polarization-Encoded Quantum Key Distribution Systems}}, 
author = {Yu, Shengrong Timothy and Yap, Jiun Yan and Liu, Mao Tong and Wang, Wenhan and Lim, Han Chuen}, 
journal = {2015 11th Conference on Lasers and Electro-Optics Pacific Rim (CLEO-PR)}, 
doi = {10.1109/cleopr.2015.7376021}, 
pages = {1--2}, 
volume = {2},
city = {Busan, Republic of Korea},
publisher = {IEEE}
}

@article{PMD_TDM5, 
year = {2018}, 
title = {{An entanglement-based wavelength-multiplexed quantum communication network}}, 
author = {Wengerowsky, S\"oren and Joshi, Siddarth Koduru and Steinlechner, Fabian and H\"ubel, Hannes and Ursin, Rupert}, 
journal = {Nature}, 
doi = {10.1038/s41586-018-0766-y},
pages = {225--228}, 
number = {7735}, 
volume = {564}
}

@article{PMD_TDM6, 
year = {2020}, 
title = {{Simple quantum key distribution with qubit-based synchronization and a self-compensating polarization encoder}}, 
author = {Agnesi, Costantino and Avesani, Marco and Calderaro, Luca and Stanco, Andrea and Foletto, Giulio and Zahidy, Mujtaba and Scriminich, Alessia and Vedovato, Francesco and Vallone, Giuseppe and Villoresi, Paolo}, 
journal = {Optica}, 
doi = {10.1364/optica.381013}, 
pages = {284}, 
number = {4}, 
volume = {7}
}

@article{PMD_TDM7, 
year = {2024}, 
title = {{Real-time polarization compensation method in quantum communication based on channel Muller parameters detection}}, 
author = {Tan, Yongjian and Wang, Jianyu and Wu, Jincai and He, Zhiping}, 
journal = {Communications Engineering}, 
doi = {10.1038/s44172-024-00198-0}, 
pages = {57}, 
number = {1}, 
volume = {3}
}

@article{PMD_TDM8, 
year = {2009}, 
title = {{Stable quantum key distribution with active polarization control based on time-division multiplexing}}, 
author = {Chen, J and Wu, G and Xu, L and Gu, X and Wu, E and Zeng, H}, 
journal = {New Journal of Physics}, 
doi = {10.1088/1367-2630/11/6/065004}, 
pages = {065004}, 
number = {6}, 
volume = {11}
}

@article{PMD_TDM9, 
    author={Wang, Jing and Rollick, Brian J. and Jia, Zhensheng and Huberman, Bernardo A.},
    journal={Journal of Lightwave Technology}, 
    title={{Time-Interleaved C-Band Co-Propagation of Quantum and Classical Channels}}, 
    year={2024},
    volume={42},
    number={11},
    pages={4086--4095},
    doi={10.1109/JLT.2024.3381105}
}

@article{PMD_WDM1, 
year = {2007}, 
title = {{Polarization drift control in fibers for entangled polarization-encoded qubits}}, 
author = {Schrenk, B. and Fedrizzi, A. and H\"ubel, H. and Poppe, A. and Zeilinger, A.}, 
journal = {2007 European Conference on Lasers and Electro-Optics and the International Quantum Electronics Conference}, 
doi = {10.1109/cleoe-iqec.2007.4386768}, 
city = {Munich, Germany},
publisher = {IEEE}
}

@article{PMD_WDM2, 
year = {2008}, 
title = {{Full polarization control for fiber optical quantum communication systems using polarization encoding}}, 
author = {Xavier, G B and Faria, G Vilela de and Tempor\~ao, G P and Weid, J P von der}, 
journal = {Optics Express}, 
doi = {10.1364/oe.16.001867}, 
pages = {1867}, 
number = {3}, 
volume = {16}
}

@article{PMD_WDM3, 
year = {2008}, 
title = {{Polarisation control schemes for fibre-optics quantum communications using polarisation encoding}}, 
author = {Faria, G Vilela de and Ferreira, J and Xavier, G B and Tempor\~ao, G P and Weid, J P von der}, 
journal = {Electronics Letters}, 
doi = {10.1049/el:20083122}, 
pages = {228}, 
number = {3}, 
volume = {44}
}

@article{PMD_WDM4, 
year = {2009}, 
title = {{Experimental polarization encoded quantum key distribution over optical fibres with real-time continuous birefringence compensation}}, 
author = {Xavier, G B and Walenta, N and Faria, G Vilela de and Tempor\~ao, G P and Gisin, N and Zbinden, H and Weid, J P von der}, 
journal = {New Journal of Physics}, 
doi = {10.1088/1367-2630/11/4/045015}, 
pages = {045015}, 
number = {4}, 
volume = {11}
}

@article{CircularBiref,
    author = {Tentori, D. and Garcia-Weidner, A.},
    year = {2013},
    title = {{Jones birefringence in twisted single-mode optical fibers}},
    journal = {Optics Express},
    volume = {21},
    number = {26},
    pages = {31725},
    doi = {10.1364/oe.21.031725}
}

@article{CircularBiref2,
    author = {Tentori, D. and Garcia-Weidner, A. and Ayala-D\'iaz, C.},
    year = {2012},
    title = {{Birefringence matrix for a twisted single-mode fiber: Photoelastic and geometrical contributions}},
    journal = {Optical Fiber Technology},
    volume = {18},
    number = {1},
    pages = {14--20},
    doi = {10.1016/j.yofte.2011.10.001}
}

@article{QunnectEntanglement,
  title = {{Automated Distribution of Polarization-Entangled Photons Using Deployed New York City Fibers}},
  author = {Craddock, Alexander N. and Lazenby, Anne and Portmann, Gabriel Bello and Sekelsky, Rourke and Flament, Mael and Namazi, Mehdi},
  journal = {PRX Quantum},
  volume = {5},
  issue = {3},
  pages = {030330},
  numpages = {7},
  year = {2024},
  doi = {10.1103/PRXQuantum.5.030330}
}

@article{ScatteringFiberSensorReviewAPL,
    author = {Lu, Ping and Lalam, Nageswara and Badar, Mudabbir and Liu, Bo and Chorpening, Benjamin T. and Buric, Michael P. and Ohodnicki, Paul R.},
    title = {Distributed optical fiber sensing: Review and perspective},
    journal = {Applied Physics Reviews},
    volume = {6},
    number = {4},
    pages = {041302},
    year = {2019},
    doi = {10.1063/1.5113955}
}

@article{FiberSensingReviewOld, 
    year = {2004}, 
    title = {{Overview of high performance fibre-optic sensing}}, 
    author = {Kirkendall, Clay K and Dandridge, Anthony}, 
    journal = {Journal of Physics D: Applied Physics}, 
    doi = {10.1088/0022-3727/37/18/r01}, 
    pages = {R197}, 
    number = {18}, 
    volume = {37}
}

@article{FiberInterferometers, 
    year = {2022}, 
    title = {{Limits and prospects for long-baseline optical fiber interferometry}}, 
    author = {Hilweg, Christopher and Shadmany, Danial and Walther, Philip and Mavalvala, Nergis and Sudhir, Vivishek}, 
    journal = {Optica}, 
    doi = {10.1364/optica.470430},  
    pages = {1238}, 
    number = {11}, 
    volume = {9}
}

@article{FiberTempSensors, 
    year = {2023}, 
    title = {{Optical Fiber Based Temperature Sensors: A Review}}, 
    author = {Gangwar, Rahul Kumar and Kumari, Sneha and Pathak, Akhilesh Kumar and Gutlapalli, Sai Dheeraj and Meena, Mahesh Chand}, 
    journal = {Optics}, 
    doi = {10.3390/opt4010013}, 
    pages = {171--197}, 
    number = {1}, 
    volume = {4}
}

@article{FiberGyroscope, 
    year = {1997}, 
    title = {{Fundamentals of the interferometric fiber-optic gyroscope}}, 
    author = {Lef\`evre, Herv\'e C.}, 
    journal = {Optical Review}, 
    doi = {10.1007/bf02935984}, 
    pages = {A20}, 
    number = {1}, 
    volume = {4}
}

@article{PolFadingReview, 
    year = {2022}, 
    title = {{Polarization Fading Suppression for Optical Fiber Sensing: A Review}}, 
    author = {Xiao, Lin and Wang, Yu and Li, Yan and Bai, Qing and Liu, Xin and Jin, Baoquan}, 
    journal = {IEEE Sensors Journal}, 
    doi = {10.1109/jsen.2022.3161075}, 
    pages = {8295--8312}, 
    number = {9}, 
    volume = {22}
}

@ARTICLE{FBGSensorReview,
  author={Caucheteur, Christophe and Guo, Tuan and Albert, Jacques},
  journal={Journal of Lightwave Technology}, 
  title={Polarization-Assisted Fiber Bragg Grating Sensors: Tutorial and Review}, 
  year={2017},
  volume={35},
  number={16},
  pages={3311-3322},
  doi={10.1109/JLT.2016.2585738}
}

@INPROCEEDINGS{FBGSensor1,
  author={Barot, Dipen and Wang, Gang and Duan, Lingze},
  booktitle={2019 IEEE Photonics Conference (IPC)}, 
  title={High Resolution Dynamic Strain Sensor using a Polarization Maintaining Fiber Bragg Grating}, 
  year={2019},
  volume={},
  number={},
  pages={1-4},
  doi={10.1109/IPCon.2019.8908376},
  city = {San Antonio, TX, USA},
  publisher = {IEEE}
}

@article{FBGSensor_PMDBad,
    author = {Todd, Michael D and Nichols, Jonathan M and Trickey, Stephen T and Seaver, Mark and Nichols, Christy J and Virgin, Lawrence N},
    year = {2007},
    title = {{Bragg grating-based fibre optic sensors in structural health monitoring}},
    journal = {Philosophical Transactions of the Royal Society A},
    volume = {365},
    pages = {317-–343},
    doi = {10.1098/rsta.2006.1937}
}

@article{Mecozzi,
    title = {{Sensing with submarine optical cables}},
    author = {Antonio Mecozzi},
    journal = {APL Photonics},
    year = {2024},
    volume = {9},
    pages = {070902},
    doi = {10.1063/5.0210825}
}

@article{OCT_PMD,
    author = {Ellen Ziyi Zhang and Wang-Yuhl Oh and Martin L. Villiger and Liang Chen and Brett E. Bouma and Benjamin J. Vakoc},
    journal = {Optics Express},
    number = {1},
    pages = {1163--1180},
    publisher = {Optica Publishing Group},
    title = {{Numerical compensation of system polarization mode dispersion in polarization-sensitive optical coherence tomography}},
    volume = {21},
    year = {2013},
    doi = {10.1364/OE.21.001163}
}

@article{OldModel1,
    author = {dal Forno, A. O. and Paradisi, A. and Passy, R. and von der Weid, J. P.},
    year = {2000},
    title = {{Experimental and theoretical modeling of polarization-mode dispersion in single-mode fibers}},
    journal = {IEEE Photonics Technology Letters},
    volume = {12},
    number = {3},
    pages = {296-–298},
    doi = {10.1109/68.826919}
}

@article{OldModel2,
    author = {Karlsson, M.},
    year = {2001},
    title = {Probability Density Functions of the Differential Group Delay in Optical Fiber Communication Systems},
    journal = {Journal of Lightwave Technology},
    volume = {19},
    number = {3},
    pages = {324-–331},
    doi = {10.1109/50.918883}
}

@article{OldModel_TimeStochastic,
  author={Antonelli, C. and Colamarino, C. and Mecozzi, A. and Brodsky, M.},
  journal={IEEE Photonics Technology Letters}, 
  title={{A Model for Temporal Evolution of PMD}}, 
  year={2008},
  volume={20},
  number={12},
  pages={1012-1014},
  doi={10.1109/LPT.2008.923774}
}

@article{HingeOG, 
    year = {2006}, 
    title = {{Polarization Mode Dispersion of Installed Fibers}}, 
    author = {Brodsky, Misha and Frigo, Nicholas J. and Boroditsky, Misha and Tur, Moshe}, 
    journal = {Journal of Lightwave Technology}, 
    issn = {0733-8724}, 
    doi = {10.1109/jlt.2006.885781}, 
    pages = {4584--4599}, 
    number = {12}, 
    volume = {24},
}

@article{CzeglediSciRep, 
    year = {2016}, 
    title = {{Polarization Drift Channel Model for Coherent Fibre-Optic Systems}}, 
    author = {Czegledi, Cristian B. and Karlsson, Magnus and Agrell, Erik and Johannisson, Pontus}, 
    journal = {Scientific Reports}, 
    doi = {10.1038/srep21217}, 
    pages = {21217}, 
    number = {1}, 
    volume = {6}
}

@article{CzeglediOFC, 
    year = {2017}, 
    title = {{Temporal Stochastic Channel Model for Absolute Polarization State and Polarization-Mode Dispersion}}, 
    author = {Czegledi, Cristian B and Karlsson, Magnus and Johannisson, Pontus and Agrell, Erik}, 
    journal = {Optical Fiber Communication Conference}, 
    doi = {10.1364/ofc.2017.th3f.2}, 
    pages = {Th3F.2}, 
    city = {Los Angeles, CA, USA},
    publisher = {Optical Society of America}
}

@book{YarivYeh,
    author = {Yariv, Amnon and Yeh, Pochi},
    title = {{Photonics: Optical Electronics in Modern Communications}},
    edition = {6},
    publisher = {Oxford University Press},
    year = {2007},
    isbn = {978-0-19-517946-0},
    city = {New York, NY, USA}
}

@book{BuckOpticalFibers,
    author = {John A. Buck},
    title = {{Fundamentals of optical fibers}},
    year = {2004},
    publisher = {Wiley},
    address = {Hoboken, NJ},
    isbn = {0471221910}
}

@book{SnyderLove,
    author = {Allan W. Snyder and John D. Love},
    title = {{Optical Waveguide Theory}},
    year = {1983},
    publisher = {Chapman and Hall},
    address = {New York, NY},
    isbn = {0412242508}
}

@article{Gloge,
    author = {Gloge, D},
    year = {1971},
    title = {{Weakly Guiding Fibers}},
    journal = {Applied Optics},
    volume = {10},
    number = {10},
    pages = {2252}, 
    doi = {10.1364/ao.10.002252}
}

@article{FusedSilica_ThreePole_Temp2,
    author = {Douglas B. Leviton and Bradley J. Frey},
    title = {{Temperature-dependent absolute refractive index measurements of synthetic fused silica}},
    journal = {Proceedings of SPIE 6273, Optomechanical Technologies for Astronomy},
    pages = {62732K},
    year = {2006},
    doi = {10.1117/12.672853},
    city = {Orlando, FL, USA},
    publisher = {SPIE}
}

@article{FusedSilica_Sellmeier_ThreePole,
    author = {I. H. Malitson},
    title = {{Interspecimen Comparison of the Refractive Index of Fused Silica}},
    year = {1965},
    volume = {5},
    number = {10},
    journal = {Journal of the Optical Society of America},
    pages = {1205--1209}
}

@article{FusedSilica_ThreePole_Temp1,
    author = {Matsuoka, J. and Kitamura, N. and Fujinaga, S. and Kitaoka, T. and Yamashita, H},
    title = {Temperature dependence of refractive index of SiO2 glass},
    year = {1991},
    journal = {Journal of Non-Crystalline Solids},
    volume = {135},
    number = {1},
    pages = {89--89},
    doi = {10.1016/0022-3093(91)90447-e}
}

@article{FusedSilica_TempDependence1,
    author = {Paulo S. Andre and Armando N. Pinto},
    title = {{Chromatic dispersion fluctuations in optical fibers due to temperature and its effects in high-speed optical communication systems}},
    journal = {Optics Communications},
    volume = {246},
    year = {2005},
    pages = {303-–311}
}

@article{FusedSilica_TwoTerm_Temp,
    title = {Temperature-Dependent Sellmeier Coefficients and Chromatic Dispersions for Some Optical Fiber Glasses},
    author = {Gorachand Ghosh and Michiyuki Endo and Takashi Iwasalu},
    journal = {Journal of Lightwave Technology},
    volume = {12},
    number = {8},
    year = {1994}
}

@article{Rashleigh, 
    year = {1983}, 
    title = {{Origins and control of polarization effects in single-mode fibers}}, 
    author = {Rashleigh, S.}, 
    journal = {Journal of Lightwave Technology}, 
    issn = {0733-8724}, 
    doi = {10.1109/jlt.1983.1072121}, 
    pages = {312--331}, 
    number = {2}, 
    volume = {1}
}

@article{OG_PMD_Analysis,
    author = {Gupta, D. and Kumar, A. and Thyagarajan, K},
    year = {2006},
    title = {{Polarization mode dispersion in single mode optical fibers due to core-ellipticity}},
    journal = {Optics Communications}, volume = {263},
    number = {1},
    pages = {36-–41},
    doi = {10.1016/j.optcom.2006.01.017}
}

@article{Mabrouki_FiberModeling,
    author = {Mabrouki, A. and Gadonna, M. and Gouronnec, A. and Goarin, R. and Naour, R. L},
    year = {1998},
    title = {{Analysis of polarization mode dispersion of single mode elliptic-core optical fibers}},
    journal = {Optics Communications},
    volume = {149},
    pages = {255-–260},
    doi = {10.1016/s0030-4018(98)00004-2}
}

@article{Boudrioua_FiberModeling,
    author = {Boudrioua, N. and Boudrioua, A. and Monteiro, F. and Losson, E. and Dandache, A. and Kremer, R.},
    year = {2008},
    title = {{Analysis of polarization mode dispersion fluctuations in single mode fibres due to temperature}},
    journal = {Optics Communications},
    volume = {281},
    number = {19},
    pages = {4870-–4875},
    doi = {10.1016/j.optcom.2008.06.005}
}

@article{ThermalStressBirefringence,
    author = {Eickhoff, W},
    year = {1982},
    title = {{Stress-induced single-polarization single-mode fiber}},
    journal = {Optics Letters},
    volume = {7},
    number = {12},
    pages = {629},
    doi = {10.1364/ol.7.000629}
}

@article{TensionBendingBirefringence,
    author = {Rashleigh, S. C. and Ulrich, R.},
    year = {1980},
    title = {{High birefringence in tension-coiled single-mode fibers}},
    journal = {Optics Letters},
    volume = {5},
    number = {8},
    pages = {354},
    doi = {10.1364/ol.5.000354}
}

@article{MMM_OG,
    author = {Jopson, R. M. and Nelson, L. E. and Kogelnik, H.},
    year = {1999},
    title = {{Measurement of second-order polarization-mode dispersion vectors in optical fibers}},
    journal = {IEEE Photonics Technology Letters},
    volume = {11},
    number = {9},
    pages = {1153-–1155},
    doi = {10.1109/68.784234}
}

@article{PMD_PNASReview, 
    year = {2000}, 
    title = {{PMD fundamentals: Polarization mode dispersion in optical fibers}}, 
    author = {Gordon, J. P. and Kogelnik, H.}, 
    journal = {Proceedings of the National Academy of Sciences}, 
    doi = {10.1073/pnas.97.9.4541}, 
    pages = {4541--4550}, 
    number = {9}, 
    volume = {97}
}

@article{Heffner,
    author = {Heffner, B. L.},
    year = {1992},
    title = {{Automated measurement of polarization mode dispersion using Jones matrix eigenanalysis}},
    journal = {IEEE Photonics Technology Letters},
    volume = {4},
    number = {9},
    pages = {1066-–1069},
    doi = {10.1109/68.157151}
}

@article{GeNVals,
    author = {James W. Fleming},
    journal = {Applied Optics},
    number = {24},
    pages = {4486--4493},
    title = {Dispersion in GeO2--SiO2 glasses},
    volume = {23},
    year = {1984},
    doi = {10.1364/AO.23.004486}
}

@article{GeNVals2,
    author = {Y. K. Prajapati and V. K. Srivastava and Vivek Singh and J. P. Saini},
    journal = {Optik},
    pages = {149--156},
    title = {Effect of germanium doping on the performance of silica based photonic crystal fiber},
    volume = {155},
    year = {2018},
    doi = {10.1016/j.ijleo.2017.10.178}
}

@article{TempDependenceSiGe,
    author = {Rego, Gaspar Mendes},
    year = {2023},
    title = {{Temperature Dependence of the Thermo-Optic Coefficient of GeO2-Doped Silica Glass Fiber}},
    journal = {Sensors},
    volume = {24},
    number = {15},
    pages = {4857},
    doi = {https://doi.org/10.3390/s24154857}
}

@article{CTEVals,
    author = {Napiorkowski, M. and Urbanczyk, W},
    year = {2023},
    title = {{Effect of core ellipticity and core-induced thermal stress on the conversion of LP 11 modes to vector vortex modes in gradually twisted highly birefringent fibers}},
    journal = {Optics Express},
    volume = {31},
    number = {6},
    pages = {9631},
    doi = {10.1364/oe.479219}
}

@article{MiscFiberProps1,
    author = {Urbanczyk, W. and Martynkien, T. and Bock, W. J},
    year = {2001},
    title = {{Dispersion effects in elliptical-core highly birefringent fibers}},
    journal = {Applied Optics},
    volume = {40},
    number = {12},
    pages = {1911},
    doi = {10.1364/ao.40.001911}
}

@article{MiscFiberProps2,
    author = {Dianov, E. M. and Mashinsky, V. M},
    year = {2005},
    title = {{Germania-Based Core Optical Fibers}},
    journal = {Journal of Lightwave Technology},
    volume = {23},
    number = {11},
    pages = {3500-–3508},
    doi = {10.1109/jlt.2005.855867}
}

@article{StressOptic_LambdaTempDependence,
    author = {Barlow, A. and Payne, D.},
    year = {1983},
    title = {{The stress-optic effect in optical fibers}},
    journal = {IEEE Journal of Quantum Electronics},
    volume = {19},
    number = {5},
    pages = {834-–839},
    doi = {10.1109/jqe.1983.1071934}
}

@article{PMDMeas1, 
    year = {2000}, 
    title = {{Long-term measurement of PMD and polarization drift in installed fibers}}, 
    author = {Karlsson, Magnus and Brentel, Jonas and Andrekson, Peter A.}, 
    journal = {Journal of Lightwave Technology}, 
    doi = {10.1109/50.850739}, 
    pages = {941--951}, 
    number = {7}, 
    volume = {18}
}

@article{PMDMeas2, 
    year = {2004}, 
    title = {{Polarization-mode dispersion of installed recent vintage fiber as a parametric function of temperature}}, 
    author = {Brodsky, M. and Magill, P. and Frigo, N.J.}, 
    journal = {IEEE Photonics Technology Letters}, 
    doi = {10.1109/lpt.2003.820113}, 
    pages = {209--211}, 
    number = {1}, 
    volume = {16}
}

@article{PMDMeas3, 
    year = {2005}, 
    title = {{Polarization Dynamics in Installed Fiberoptic Systems}}, 
    author = {Boroditsky, Misha and Brodsky, Misha and Frigo, Nicholas J. and Magil, Peter and Rosenfeldt, Harald}, 
    journal = {2005 IEEE LEOS Annual Meeting Conference Proceedings}, 
    doi = {10.1109/leos.2005.1548054}, 
    pages = {414--415},
    city = {Sydney, NSW, Australia},
    publisher = {IEEE}
}

@inproceedings{PyPol,
  author       = {J. Hoyo and L. M. Sanchez-Brea and A. Soria-Garcia},
  title        = {{Open source library for polarimetric calculations ``py\_pol''}},
  year         = {2021},
  booktitle    = {Proceedings of SPIE 11875: Computational Optics},
  organization    = {SPIE},
  pages        = {1187506},
  doi = {10.1117/12.2597163},
  city = {Online only}
}

@article{Manufacturing_RefractiveIndex,
    author = {Butov, O. V. and Golant, K. M. and Tomashuk, A. L. and van Stralen, M. J. N. and Breuls, A. H. E.},
    year = {2002},
    title = {{Refractive index dispersion of doped silica for fiber optics.}},
    journal = {Optics Communications},
    volume = {213},
    pages = {301-–308},
    doi = {10.1016/s0030-4018(02)02087-4}
}

@article{OG_POTDR,
    author = {A. J. Rogers},
    journal = {Applied Optics},
    number = {6},
    pages = {1060--1074},
    title = {{Polarization-optical time domain reflectometry: a technique for the measurement of field distributions}},
    volume = {20},
    year = {1981},
    doi = {10.1364/AO.20.001060}
}

@article{POTDR_1999,
    author = {B. Gisin},
    journal = {Journal of Lightwave Technology},
    number = {10},
    pages = {1843},
    title = {{Distributed PMD Measurement with a Polarization-OTDR in Optical Fibers}},
    volume = {17},
    year = {1999}
}

@article{POTDR_2022,
    author = {Jiahao Huo and Jian Wang and Yaping Wang and Chao Shang and Wei Huangfu and Lanlan Liu and Zhi Wang and Keping Long and Chongqing Wu},
    journal = {Applied Optics},
    number = {13},
    pages = {3754--3760},
    title = {{Vector distribution measurement of PMD in optical fiber links employing a wavelength-tunable SOP-OTDR}},
    volume = {61},
    year = {2022},
    doi = {10.1364/AO.454671}
}

@article{Validation_Cameron,
    author = {Cameron, J. and Chen, L. and Bao, X. and Stears, J.},
    year = {1998},
    title = {{Time evolution of polarization mode dispersion in optical fibers}},
    journal = {IEEE Photonics Technology Letters},
    volume = {10},
    number = {9},
    pages = {1265-–1267},
    doi = {10.1109/68.705611}
}

@book{RecentProgress,
    editor = {Yasin, Moh. and Harun, Sulaiman W. and Hamzah, Arof},
    title = {{Recent Progress in Optical Fiber Research}},
    publisher = {IntechOpen},
    isbn = {9789535149347},
    year = {2012},
    city = {Rijeka, Croatia}
}

@book{Agrawal,
    author = {Govind P. Agrawal},
    title = {{Nonlinear fiber optics}},
    publisher = {Academic Press},
    isbn = {9780123973078},
    year = {2013},
    edition = {5},
    city = {Burlington, MA, USA}
}

@article{ESD_PMD1, 
    year = {2010}, 
    title = {{Loss of polarization entanglement in a fiber-optic system with polarization mode dispersion in one optical path}}, 
    author = {Brodsky, Misha and George, Elizabeth C and Antonelli, Cristian and Shtaif, Mark}, 
    journal = {Optics Letters}, 
    issn = {0146-9592}, 
    doi = {10.1364/ol.36.000043},
    pages = {43}, 
    number = {1}, 
    volume = {36}
}

@article{ESD_PMD2, 
    year = {2011}, 
    title = {{Sudden Death of Entanglement Induced by Polarization Mode Dispersion}}, 
    author = {Antonelli, Cristian and Shtaif, Mark and Brodsky, Misha}, 
    journal = {Physical Review Letters}, 
    issn = {0031-9007}, 
    doi = {10.1103/physrevlett.106.080404},
    pages = {080404}, 
    number = {8}, 
    volume = {106}
}

@article{ESD_PMD3, 
    year = {2011}, 
    title = {{Nonlocal compensation of polarization mode dispersion in the transmission of polarization entangled photons}}, 
    author = {Shtaif, Mark and Antonelli, Cristian and Brodsky, Misha}, 
    journal = {Optics Express}, 
    doi = {10.1364/oe.19.001728}, 
    pmid = {21368986}, 
    eprint = {1012.2013},
    pages = {1728}, 
    number = {3}, 
    volume = {19}
}

@article{ESD_PDL, 
    year = {2018}, 
    title = {{Effect of Polarization Dependent Loss on the Quality of Transmitted Polarization Entanglement}}, 
    author = {Kirby, Brian T. and Jones, Daniel E. and Brodsky, Michael}, 
    journal = {Journal of Lightwave Technology}, 
    issn = {0733-8724}, 
    doi = {10.1109/jlt.2018.2879754}, 
    pages = {95--102}, 
    number = {1}, 
    volume = {37}
}

@article{Multicore1,
  title = {{Multidimensional Entanglement Generation with Multicore Optical Fibers}},
  author = {G\'omez, E.S. and G\'omez, S. and Machuca, I. and Cabello, A. and P\'adua, S. and Walborn, S.P. and Lima, G.},
  journal = {Physical Review Applied},
  volume = {15},
  issue = {3},
  pages = {034024},
  numpages = {10},
  year = {2021},
  doi = {10.1103/PhysRevApplied.15.034024}
}

@article{Multicore2,
  title = {{High-dimensional decoy-state quantum key distribution over multicore telecommunication fibers}},
  author = {Ca\~nas, G. and Vera, N. and Cari\~ne, J. and Gonz\'alez, P. and Cardenas, J. and Connolly, P. W. R. and Przysiezna, A. and G\'omez, E. S. and Figueroa, M. and Vallone, G. and Villoresi, P. and da Silva, T. Ferreira and Xavier, G. B. and Lima, G.},
  journal = {Physical Review A},
  volume = {96},
  issue = {2},
  pages = {022317},
  numpages = {8},
  year = {2017},
  doi = {10.1103/PhysRevA.96.022317}
}

@article{Multicore3,
    author = {Davide Bacco and Nicola Biagi and Ilaria Vagniluca and Tetsuya Hayashi and Antonio Mecozzi and Cristian Antonelli and Leif K. Oxenl{\o}we and Alessandro Zavatta},
    journal = {Photonics Research},
    number = {10},
    pages = {1992--1997},
    title = {Characterization and stability measurement of deployed multicore fibers for quantum applications},
    volume = {9},
    year = {2021},
    doi = {10.1364/PRJ.425890},
}

@article{Hollowcore1,
    author = {Michael Antesberger and Carla M. D. Richter and Francesco Poletti and Radan Slav\'{i}k and Periklis Petropoulos and Hannes H\"{u}bel and Alessandro Trenti and Philip Walther and Lee A. Rozema},
    journal = {Optica Quantum},
    number = {3},
    pages = {173--180},
    title = {Distribution of telecom entangled photons through a 7.7 km antiresonant hollow-core fiber},
    volume = {2},
    year = {2024},
    doi = {10.1364/OPTICAQ.514257}
}

@article{Hollowcore2,
    author = {Umberto Nasti and Hesham Sakr and Ian A. Davidson and Francesco Poletti and Ross J. Donaldson},
    journal = {Applied Optics},
    number = {30},
    pages = {8959--8966},
    title = {Utilizing broadband wavelength-division multiplexing capabilities of hollow-core fiber for quantum communications},
    volume = {61},
    year = {2022},
    doi = {10.1364/AO.471632}
}

@article{Coex1,
    title = {{Dense wavelength multiplexing of 1550 nm QKD with strong classical channels in reconfigurable networking environments}},
    author = {N A Peters and P Toliver and T E Chapuran and R J Runser and S R McNown and C G Peterson and D Rosenberg and N Dallmann and R J Hughes and K P McCabe and J E Nordholt and K T Tyagi},
    journal = {New Journal of Physics},
    volume = {11},
    year = {2009},
    doi = {10.1088/1367-2630/11/4/045012}
}

@article{Coex2,
    title = {{Quantum key distribution and 1 Gbps data encryption over a single fibre}},
    author = {P Eraerds and N Walenta and M Legr\'e and N Gisin and H Zbinden},
    year = {2010},
    journal = {New Journal of Physics},
    volume = {12},
    pages = {063027},
    doi = {10.1088/1367-2630/12/6/063027}
}

@article{MullerSampling,
    author = {Muller, Mervin E},
    title = {{A note on a method for generating points uniformly on n-dimensional spheres}},
    journal = {Communications of the ACM 2.4},
    year = {1959},
    pages = {19--20}
}

@article{TwoTermGermaniaTempDependence,
author = {Norman P. Barnes and Martin S. Piltch},
journal = {Journal of the Optical Society of America},
number = {1},
pages = {178--180},
title = {Temperature-dependent Sellmeier coefficients and nonlinear optics average power limit for germanium},
volume = {69},
year = {1979},
doi = {10.1364/JOSA.69.000178}
}

@article{Chapman,
    author = {Joseph C. Chapman and Muneer Alshowkan and Kazi Reaz and Tian Li and Mariam Kiran}, 
    title = {Continuous automatic polarization channel stabilization from heterodyne detection of coexisting dim reference signals},
    journal = {Optics Express},
    volume = {32},
    number = {26},
    pages = {47589--47619},
    year = {2024},
    doi = {10.1364/OE.543704}
}

@ARTICLE{DT1,
  author={Faruk, Md. Saifuddin and Savory, Seb J.},
  journal={Journal of Lightwave Technology}, 
  title={Measurement Informed Models and Digital Twins for Optical Fiber Communication Systems}, 
  year={2024},
  volume={42},
  number={3},
  pages={1016-1030},
  doi={10.1109/JLT.2023.3328765}
}

@ARTICLE{DT2,
  author={Zhang, Yao and Zhang, Min and Song, Yuchen and Shi, Yan and Zhang, Chunyu and Ju, Cheng and Guo, Bingli and Huang, Shanguo and Wang, Danshi},
  journal={Journal of Optical Communications and Networking}, 
  title={Building a digital twin for large-scale and dynamic C+L-band optical networks}, 
  year={2023},
  volume={15},
  number={12},
  pages={985-998},
  doi={10.1364/JOCN.503265}
}

@ARTICLE{DT3,
  author={Borraccini, Giacomo and Straullu, Stefano and Giorgetti, Alessio and Ambrosone, Renato and Virgillito, Emanuele and D’Amico, Andrea and D’Ingillo, Rocco and Aquilino, Francesco and Nespola, Antonino and Sambo, Nicola and Cugini, Filippo and Curri, Vittorio},
  journal={IEEE Transactions on Network and Service Management}, 
  title={Experimental Demonstration of Partially Disaggregated Optical Network Control Using the Physical Layer Digital Twin}, 
  year={2023},
  volume={20},
  number={3},
  pages={2343-2355},
  doi={10.1109/TNSM.2023.3288823}
}

@ARTICLE{DTReview1,
  author={Wang, Danshi and Song, Yuchen and Zhang, Yao and Jiang, Xiaotian and Dong, Jiawei and Khan, Faisal Nadeem and Sasai, Takeo and Huang, Shanguo and Lau, Alan Pak Tao and Tornatore, Massimo and Zhang, Min},
  journal={Journal of Lightwave Technology}, 
  title={Digital Twin of Optical Networks: A Review of Recent Advances and Future Trends}, 
  year={2024},
  volume={42},
  number={12},
  pages={4233-4259},
  doi={10.1109/JLT.2024.3401419}
}

@article{DTReview2,
    author = {Yang, Hang and Niu, Zekun and Fan, Qirui and Li, Lyu and Shi, Minghui and Zeng, Chuyan and Xiao, Shilin and Hu, Weisheng and Yi, Lilin},
    title = {The Digital Twin Framework for the Physical Wideband and Long-Haul Optical Fiber Communication Systems},
    journal = {Laser \& Photonics Reviews},
    volume = {18},
    number = {10},
    pages = {2400234},
    doi = {10.1002/lpor.202400234},
    year = {2024}
}

@article{DTWireless,
    author={Iye, Tetsuya and Sakamoto, Masaya and Takaya, Shohei and Sato, Eisaku and Susukida, Yuki and Nagaoka, Yu and Maruta, Kazuki and Nakazato, Jin},
    journal={IEEE Access}, 
    title={{Open Wireless Digital Twin: End-to-End 5G Mobility Emulation With OpenAirInterface and Ray Tracing}}, 
    year={2025},
    volume={13},
    number={},
    pages={175109--175122},
    doi={10.1109/ACCESS.2025.3619105}
}

\end{document}